\documentclass[11pt]{article}

\usepackage{fullpage}
\usepackage[pdftex]{graphicx}
\usepackage{url, float, lscape, verbatim, bigints}
\usepackage{xcolor, dsfont, pdfpages, sectsty, MnSymbol}

\usepackage{amsmath} %, amsfonts,amsthm}
\usepackage{psfrag,epsf}
\usepackage{latexsym}
\usepackage{calc}
\usepackage[ragged, symbol]{footmisc}
\usepackage{grffile}
\usepackage{booktabs}
\usepackage{algorithmic}
\usepackage{algorithm}
\usepackage{caption}
\usepackage{comment}
\usepackage{color}

\usepackage[nonamebreak, sort&compress, round, authoryear]{natbib}
\usepackage{breakcites}
\usepackage{etoolbox}
\makeatletter
\patchcmd{\@citex}{,}{;}{}{}
\makeatother

\usepackage[inline]{enumitem}
\makeatletter
\newcommand{\inlineitem}[1][]{%
	\ifnum\enit@type=\tw@
	{\descriptionlabel{#1}}
	\hspace{\labelsep}
	\else
	\ifnum\enit@type=\z@
	\refstepcounter{\@listctr}\fi
	\quad\@itemlabel\hspace{\labelsep}
	\fi}
\makeatother

\newcommand\pN{\mathcal{N}}
\newcommand{\blind}{0}

\usepackage[left=1.1in, right = 1.1in, top = 1.1in, bottom = 1.1in]{geometry}
\renewcommand{\baselinestretch}{1.3}

\begin{document}

	\def\spacingset#1{\renewcommand{\baselinestretch}%
		{#1}\small\normalsize} \spacingset{1}

	%%%%%%%%%%%%%%%%%%%%%%%%%%%%%%%%%%%%%%%%%%%%%%%%%%%%%%%%%%%%%%%%%%%%%%%%%%%%%%
	
	\if0\blind
	{
		\title{\bf Longitudinal Dynamic Functional Regression}
		\author{Md Nazmul Islam\thanks{
				Department of Statistics, North Carolina State University (Email: \textit{mnislam@ncsu.edu})}
			\and
			Ana-Maria Staicu\thanks{
				Department of Statistics, North Carolina State University (Email: \textit{astaicu@ncsu.edu}; corresponding author)}
			\and
			Eric van Heugten\thanks{
				Department of Animal Science, North Carolina State University (Email: \textit{evheugte@ncsu.edu})}
			\and
}			
  \date{}
  \maketitle
} \fi
	
	\if1\blind
	{
		\bigskip
		\bigskip
		\bigskip

\begin{center}
	{\bf \Large  Longitudinal Dynamic Functional Regression}\\ 
\end{center}
 \medskip
} \fi

\baselineskip=16pt

\section*{Abstract}

%We consider regression models to study the dynamic association between scalar outcomes and functional predictors observed over time, at many instances. We propose a flexible yet parsimonious approach to model the time-varying association. The proposed method allows to reconstruct the full response trajectory. Numerical investigation through simulation studies and data analysis show a superior performance in terms of accurate prediction and efficient computations. The methods are inspired and applied to an animal science application, where of interest is to study the association between the feed intake of lactating sows and the minute-by-minute{temperature} throughout the 21st days of their lactation period. R code and an R illustration are provided at \url{http://www4.stat.ncsu.edu/~staicu/software}. 

This article develops flexible methodology to study the association between scalar outcomes and functional predictors observed over time, at many instances, in longitudinal studies. We propose a parsimonious modeling framework to study time-varying regression that leads to superior prediction properties and allows to reconstruct full trajectories of the response. The idea is to model the time-varying functional predictors using orthogonal basis functions and expand the time-varying regression coefficient using the same basis. Numerical investigation through simulation studies and data analysis show excellent performance in terms of accurate prediction and efficient computations, when compared with existing alternatives.
The methods are inspired and applied to an animal science application, where of interest is to study the association between the feed intake of lactating sows and the minute-by-minute {temperature} throughout the 21st days of their lactation period. R code and an R illustration are provided at \url{http://www4.stat.ncsu.edu/~staicu/software}.

\textbf{\underline{Keywords:}} Functional data; Functional principal component analysis; Longitudinal study; Longitudinal functional regression; Penalization; Time-varying coefficient model.

\section{Introduction}

\begin{figure}[H]\centering
	\begin{tabular}{cc}
		\hspace{-.8cm} 
		\includegraphics[angle=0, height=4cm, width=6.5cm]{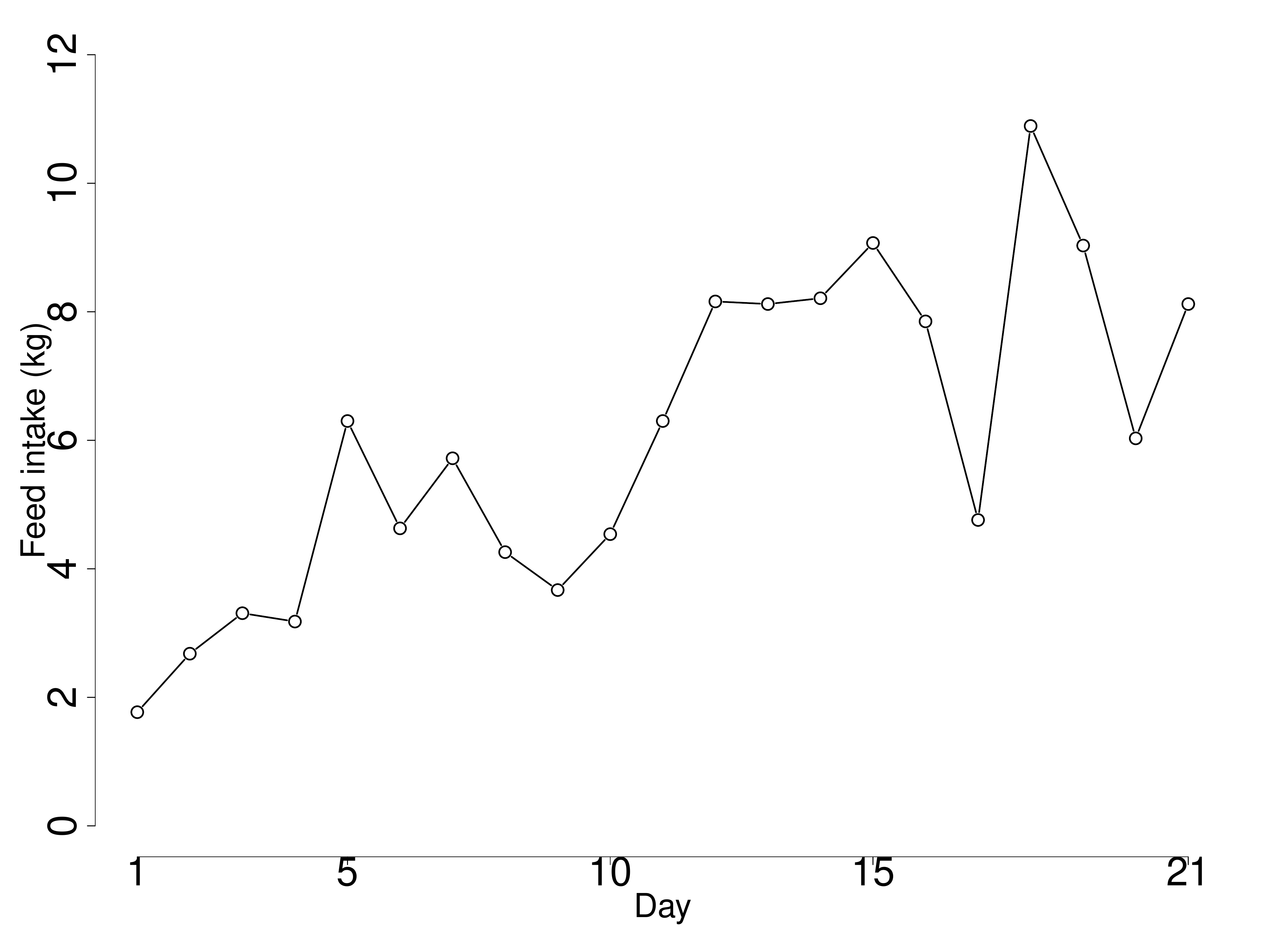}&
		\hspace{-.2cm} 
		\includegraphics[angle=0, height=4cm, width=6.5cm]{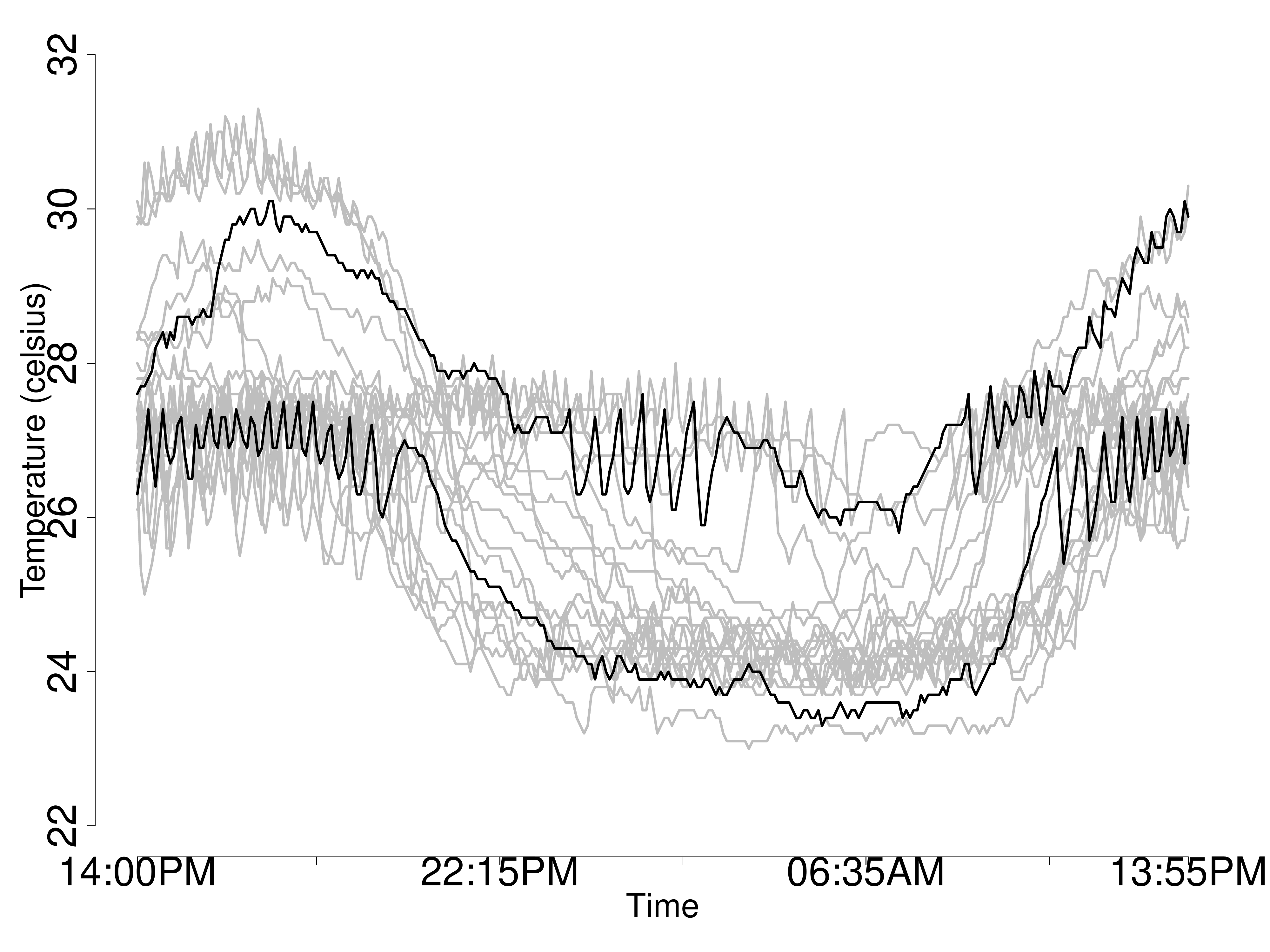}  
	\end{tabular}
	\caption{\small Data for a randomly chosen sow; feed-intake (kg) and {temperature} ($^{\circ}$C) profiles from left to right respectively.
		\label{Sow14473}}
\end{figure}

Functional regression has attracted a lot of interest in recent years; see \cite{ramssilv, fan2000two, cardot1999functional, cardot2003spline, muller2005functional,  cai2006prediction, morris2006wavelet, reiss2007functional,  ivanescu2015penalized, scheipl2015functional} to name a few. In this paper we consider longitudinal scalar-on-function regression for scalar outcomes and functional predictors observed repeatedly. This research is motivated by an animal science study of the effect of daily ambient air{temperature} on feed intake of sows during their lactation period. To be specific, a number of sows are observed for several days during their lactation period and for each day, the total daily feed intake, as well as, the minute-by-minute daily {temperature} for the respective day are recorded. Figure \ref{Sow14473} illustrates the data for a randomly chosen sow: daily feed-intake for each lactation day (left panel) and {temperature} profiles (right). 

Functional linear model (FLM) for scalar-on-function regression is a popular regression model and assumes that the effect of the functional predictor is captured by the integral of the predictor weighted by a smooth regression coefficient. Three estimation approaches are quite common: both the functional predictor and smooth coefficient are expanded using the empirical eigenbasis of the predictors covariance (\cite{cardot1999functional}); both the functional predictor and smooth coefficient are expanded using B-spline basis and penalties are employed to control the smoothness of the parameter function (\cite {ramssilv}); or a mixture of these approaches, predefined basis function is used to represent the smooth parameter, the empirical eigenbasis of the predictors covariance is used to expand the functional predictors, and in addition penalties are used to control the smoothness of the parameter function (\cite{cardot2003spline, goldsmith2011penalized}). Extensions of these approaches to accommodate additional covariates or more flexible relationships have been discussed previously by \cite{james2005functional, morris2006wavelet, cardot2008varying, muller2012functional,mclean2014functional}.

The popular scalar-on-function regression model has been extended to analyze longitudinal functional data where the scalar response and the functional predictor are observed repeatedly for each subject. \cite{goldsmith2012longitudinal} introduced longitudinal penalized functional regression (LPFR), which assumes that the effect of the functional predictor is constant over time. The authors model the time-invariant regression coefficient via the truncated polynomial basis and use a penalized-based approach for estimation. In a similar spirit, \cite{gertheiss2013longitudinal} introduced longitudinal functional principal component regression, where the response is regressed onto the functional principal components of the various structures that compose the functional predictors, which are obtained using longitudinal functional principal component analysis (\cite{greven2011longitudinal}). The main limitation of these methods is the assumption that the predictor is invariant over time. Such assumption may be viewed as strong and unrealistic in some situations. For example, in the motivating animal science study, a lactating sow's body can adjust to the prolonged exposure to heat and, thus, the effect of the heat on their feed intake behavior is expected to change throughout their lactation period. Therefore, assuming a 
time-varying relationship between the temperature and the feed intake is more appropriate.

Recently, \cite{kundu2016longitudinal} considered a time-varying functional coefficient in this longitudinal functional model framework. The authors model the bivariate regression coefficient as a linear combination of \emph{user specified} time-dependent basis functions with unknown time-invariant coefficient functions. While these time-invariant functions are estimated via a penalty operator that is informed by the scientific context of the problem, the time-dependent basis functions are chosen arbitrarily and their number is then selected using Akaike information criterion or pointwise confidence intervals for the  time-invariant coefficient functions. In simulations and data application, the authors use two temporal functions $f_1(t)=1$ and $f_2(t)=t$; their methodology is implemented solely for polynomial temporal functions. In general it is not clear how to select an optimal temporal basis. In addition, the methodology is limited to Gaussian responses and it does not seem trivial to extend it to non-Gaussian cases. Our numerical experience with this method indicated that, when applicable, the approach is rather computationally expensive. Furthermore, like LPFR and longitudinal functional principal component regression, this method too does not consider reconstruction of full response trajectory, which is often a major goal in longitudinal studies involving a repeatedly measured response.

In this paper, we propose the longitudinal dynamic functional regression (in short LDFR) for longitudinal scalar-on-function regression that accounts for a time-varying smooth effect of the functional predictors. There are three key novelties of this paper. The first novelty is the proposed model for both Gaussian and non-Gaussian longitudinal responses. Only \cite{kundu2016longitudinal} studied regression models for longitudinal functional covariates that allow a time-varying effect of the covariates, but their study is limited to Gaussian responses. The second novelty is the use of a combination of ideas from functional data analysis and longitudinal data analysis that has important advantages: 1) it allows to tackle a challenging problem that has not previously been solved in this generality; 2) it takes into account the dependence among the functional predictors; and 3) it allows prediction of the full trajectory of the response. The third novelty is the way the smoothness of the time-varying regression coefficient is modeled, by using a mixture of popular approaches in non-parametric regression. The main advantages of this approach are: (i) it enjoys a parsimonious modeling framework; (ii) it does not require any information about the temporal behavior of the functional coefficient; (iii) it provides excellent numerical performance in terms of prediction accuracy; (iv) it is computationally efficient; and (v) it can be easily implemented using well-developed freely available software.

The structure of this paper is as follows. Section \ref{sec:method} introduces the proposed methodology for responses in the exponential family. Section \ref{sec:estimation} describes the estimation procedure of the parameters of interest. Section \ref{sec:prediction} details the response trajectory prediction. Numerical assessment of the proposed method is described in Section \ref{sec:simulation} in a simulation study and in Section \ref{sec:data_application}, via our motivating application. Finally, Section \ref{sec:discussion} concludes our paper. %points out limitations of the proposed approach and discusses feasible extensions. 

\section{Proposed methodology}\label{sec:method}

Let the observed data be  $[t_{ij} , Y_{ij}, \{  ( W_{ijr}, s_{ijr}):r = 1, \ldots ,r_{ij} \}; i = 1, \cdots, I, j = 1,2,..., n_{i}];$  where $i$ indexes the subject, $j$ indexes the repeated observations, $Y_{ij}$ is the response measured at time $t_{ij},$ and  $W_{ijr}$'s are the noisy functional covariates observed at points $s_{ijr}$. It is assumed that $t_{ij} \in \mathcal{T}$ and $s_{ijr} \in \mathcal {S}$ for closed and compact sets $\mathcal{T}$ and $\mathcal{S}$, respectively. We assume that for each $i$ and $j$, $r_{ij}$ is large, and furthermore that the set $\{s_{ij1}, \cdots, s_{ijr_{ij}}\}$ is dense in $\mathcal{S}$. Also we assume that, while $n_i$ may be small for all $i$, the set $\{t_{ij} : j=1, \ldots, n_i, i=1, \ldots, I\}$ is dense in $\mathcal{T}$. We consider that $W_{ijr} = X_{ij}(s_{ijr}) + \epsilon^{w}_{ij}(s_r),$ where $X_{ij}(\cdot)$ is the latent functional predictor corresponding to the subject $i$ and time $t_{ij}$, and $\epsilon^{w}_{ij}(\cdot)$ is a zero-mean measurement error process.

Our objective is to develop association models for time-varying responses $Y_{ij}=Y_i(t_{ij})$ and true smooth time-varying functional covariate $X_{ij}(\cdot)=X_i(\cdot, t_{ij})$.  Specifically, we consider longitudinal dynamic functional regression models that account for time-varying smooth effect:
\begin{equation}
\label{eqn:LFR_model}  
\begin{aligned}
Y_{ij} \lvert X_{ij}(\cdot) &\sim \text {EF}( \mu_{ij} , \eta ),\\
g( \mu_{ij} ) &= \alpha(t_{ij}) + \int_{\mathcal{S}} X_{ij}(s)\gamma(s,t_{ij})ds +Z_{b,ij} b_{i} ,\\
\end{aligned}
\end{equation}
where $ EF( \mu_{ij}  , \eta ) $ denotes the exponential family with mean $ \mu_{ij} =\mu_i(t_{ij})$ and dispersion parameter $ \eta $ and $ g(\cdot) $ is a known, monotone link function. Here, $\alpha(\cdot)$ is an unknown intercept function, and $\gamma(\cdot, t_{ij})$ is an unknown regression coefficient function that quantifies the time-varying effect of $X_{ij}(\cdot)$ on the conditional mean response of $Y_{ij}$ at time $t_{ij}$ and is the main object of interest. Also $b_{i}$ is a subject specific $q$-dimensional vector and $Z_{b, ij}$ is its associated $1\times q$-dimensional random design matrix.  It is assumed that $b_i$ is independent and identically distributed (IID) as $N_q(0_q, D)$ where $0_q$ is the $q$-dimensional vector of zeros and $D$ is $q\times q$ covariance matrix. Both the intercept function $\alpha(\cdot)$ and the regression coefficient function $\gamma(\cdot, \cdot)$ are assumed to be smooth. Remark that, just like the functional linear model, the regression coefficient $\gamma(\cdot, t)$ is only identifiable up to the space spanned by the $X_{ij}(\cdot)$'s. Equivalently, if $h(\cdot)$ is in the orthogonal complement of this space then both $\gamma(s, \cdot)$ and $\gamma(s, \cdot) + h(s)$ yield the same association with the response; nevertheless the integral term is identifiable. Model (\ref{eqn:LFR_model}) has been introduced in \cite{kundu2016longitudinal}. The key difference is that \cite{kundu2016longitudinal} assume that $\gamma(s,\cdot)$ is a linear combination of known, but arbitrarily chosen, temporal functions; for example $\gamma(s, t)=\gamma_0(s) +t\gamma_1(s) +t^2\gamma_2(s)$ with unknown functions $\gamma_0(\cdot), \gamma_1(\cdot)$, and $\gamma_2(\cdot)$. We impose no such limitation and propose a parsimonious modeling framework that results in very competitive prediction performance and computational time. For convenience, assume $s_{ijr}=s_r$.

Our modeling approach consists of two parts. 
First, let $\{\phi_{k} (\cdot) : k \geq 1 \}$ be a time-invariant orthonormal basis in $L^{2}(\mathcal{S})$; i.e.  $ \int_{\mathcal{S}} \phi_{k} (s) \phi_{k'} (s) ds = 1 $ if $ k = k' $ and 0 otherwise. Consider the expansion of the functional regression coefficient $\gamma(\cdot, t_{ij})$ using this same  basis function $\gamma(s,t)  =  \sum_{k=1}^{\infty} \beta_{k}(t) \phi_{k} (s)$, where $\beta_{k}(\cdot)$ is an unknown smooth function of time and defined uniquely by  $\beta_{k}(t) = \int_{\mathcal{S}} \gamma(s,t) \phi_{k} (s) ds $. Often, the infinite summation is truncated at some finite level, say $K$ so that $\gamma(s,t)  \approx  \sum_{k=1}^{K} \beta_{k}(t) \phi_{k} (s)$. In some sense the truncation $K$ quantifies the smoothness of the function $\gamma(s,t)$ in the $s$-direction: a small value of $K$ results in over smooth function in this direction, while a large value gives wiggly behavior in this direction. Assuming that $\gamma (\cdot, \cdot)$ is a second order differentiable function, we can describe the smoothness in the $t$-direction by the total curvature in this direction, $\|\partial^2 \gamma(\cdot, \cdot)/\partial t^2\|^2 = \int \int \{\partial^2 \gamma(s,t)/\partial t^2 \} ds dt$. A direct extension to the popular non-parametric estimation approach of an unknown smooth is to control the amount of smoothness in $t$, that is to estimate  
$\gamma(\cdot, \cdot)$ via the minimization of a penalized criterion
%\begin{eqnarray}
%\label{eqn:penalized_criterion_general}
$
-2\sum_{i=1}^I\sum_{j=1}^{n_i} \log f(Y_{ij} |b_i) -2 \sum_{i=1}^I\log f_b(b_i) + \lambda_\alpha \|\alpha^{''}(\cdot)\|^2  + \lambda \|\partial^2 \gamma(\cdot, \cdot)/\partial t^2\|^2 ,
$ %\end{eqnarray}
where $f(Y_{ij}|b_i)$ is the density specified by the exponential family model (\ref{eqn:LFR_model}) and $f_b(b_i)$ is the density of the random terms. Typically $f_b(b) = \exp(-b^T D^{-1}b/2)$, corresponding to a multivariate normal with mean vector $0$ and variance-covariance matrix $D$, and by ignoring the multiplicative term $\{det(2\pi D)\}^{-1/2}$ where $det(D)$ is the determinant of $D$. Here $\lambda_\alpha>0$ is a smoothing parameter that controls the amount of smoothing of the unknown function $\alpha(\cdot)$ and $\lambda>0$ is an unknown parameter that controls the smoothness of  $\gamma(\cdot, \cdot)$ in direction $t$, relative to the goodness of fit, quantified by the likelihood term. Using the representation of $\gamma(\cdot, \cdot)$ via $K$ orthogonal basis functions $\phi_k(\cdot)$'s, the penalized criterion is approximated by
\begin{eqnarray}
\label{eqn:penalized_criterion_general}
-2\sum_{i=1}^I\sum_{j=1}^{n_i} \log f(Y_{ij} |b_i) -2 \sum_{i=1}^I\log f_b(b_i) + \lambda_\alpha \|\alpha^{''}(\cdot)\|^2  + \lambda \sum_{k=1}^K \|\beta^{''}_k(\cdot) \|^2 ,
\end{eqnarray}
as $\|\partial^2 \gamma(\cdot, \cdot)/\partial t^2\|^2 = \sum_{k=1}^K \|\beta^{''}_k(\cdot) \|^2$ due to the orthogonality of the basis $\phi_k(\cdot)$'s. The penalized criterion (\ref{eqn:penalized_criterion_general})
essentially says that we control the smoothness of the unknown bivariate function  $\gamma(\cdot, \cdot)$ through two parameters - $K$ and $\lambda$ - but in a distinct way from the common practice \citep{ruppert2003semiparametric, wood2006generalized, wood2006confidence}. Combining different approaches to control the smoothness of a multivariate function is inspired from \cite{kim2017additive} in the context of function-on-function regression. 

Using the orthogonal time-invariant basis $\{\phi_k(\cdot)\}$'s we represent the longitudinal functional covariates as $X_{ij}(s)  = \sum_{k=1}^{\infty} \xi_{ijk} \phi_{k}(s) $, where $ \xi_{ijk} = \int_{\mathcal{S}} X_{ij}(s) \phi_{k}(s) ds $. The time-varying basis coefficients $\xi_{ijk}$ are dependent over $j$, due to the dependence of the functional covariates within the same subject. It follows that $\int_{\mathcal{S}} X_{ij}(s)\gamma(s,t_{ij})ds = \sum_{k=1}^{\infty} \xi_{ijk} \beta_{k}(t_{ij})$, which yields the following more convenient representation of the model (\ref{eqn:LFR_model}): $
g(\mu_{ij} )= \alpha(t_{ij}) + \sum_{k=1}^{\infty} \xi_{ijk} \beta_{k}(t_{ij})+Z_{b, ij} b_i$. Corresponding to the truncation $K$, the model (\ref{eqn:LFR_model}) is approximated by:
\begin{eqnarray}
g(\mu_{ij})= \alpha (t_{ij}) +  \sum_{k=1}^{ K }   \xi_{ijk} \beta_k (t_{ij}) +Z_{b,ij} b_i, \label{eqn:integral:finite_summation_expansion1}
\end{eqnarray} 
which is a well researched model in the longitudinal literature, if $\xi_{ijk}$ were known. The model parameters of (\ref{eqn:integral:finite_summation_expansion1}) can be estimated using the penalized criterion (\ref{eqn:penalized_criterion_general}), which assumes that the coefficient functions $\{\beta_k(\cdot)\}$'s have all the same type of smoothness. 

Incorporating additional covariates via the modeling framework (\ref{eqn:LFR_model}) carries on in a straightforward manner to (\ref{eqn:penalized_criterion_general}) and (\ref{eqn:integral:finite_summation_expansion1}), irrespective whether the covariates are modeled using a linear or a smooth dependence. There are two key challenges in this approach: 1) selection of the orthogonal basis $\{ \phi_k(\cdot)\}_{k\geq 1}$ as this directly impacts the selection of the truncation $K$ and 2) the estimation of the basis coefficients $\xi_{ijk}$ from the observed noisy functional covariates $W_{ij}(\cdot)$.

\subsection{Selection of the orthogonal basis}

There are several approaches to select $\phi_{k}(\cdot)$'s. One approach is using a pre-specified basis similar to \cite{zhou2008joint}. Another approach is using the eigenbasis of some appropriately chosen covariance function; see \cite{park2015longitudinal}. Specifically we consider the marginal covariance function induced by the obseved functional covariates and select the basis as the eigenbasis of this covariance function.  %The approach also takes into account the fact that $X_{ij}(\cdot)$'s are observed indirectly through the noisy functional covariates $W_{ijr}=W_{ij}(s_r)$.

Recall that the functional covariate is viewed as the sum of two independent processes $W_{ij}(s) = X_{ij}(s) + \epsilon_{ij}^{w}(s)$, where by an abuse of notation we write $W_{ij}(s_r) = W_{ijr}$ and $X_{ij}(\cdot) = X_{i}(\cdot, t_{ij}).$ We assume that $X_{i}(\cdot, \cdot)$'s are IID over $i,$ and comprise the subject-specific deviation; in contrast, $\epsilon^{w}_{ij}(\cdot)$'s are IID over $i$ and $j,$ and characterize the time-specific deviation from the subject-specific trend. Furthermore both $X_i(\cdot, \cdot)$ and $\epsilon_{ij}^{w}(\cdot)$ are zero-mean processes. Define $\Sigma( s, s') = \int_{\mathcal{T}} E[ X_{i}(s, t) X_{i}(s', t)] h(t) dt$, where $h(\cdot)$ is the sampling density of the time points $t_{ij}$'s; see \cite{park2015longitudinal} for justification that this function is a proper covariance function (positive semidefinite and symmetric function). We call this the "marginal covariance function" induced by the latent signal $X_{i}.$ Assume the covariance of the error process can be written as the sum between a smooth covariance function and a nugget effect such as $cov(\epsilon_{ij}^{w}(s),\epsilon_{ij}^{w}(s')) = \Gamma(s,s') + \sigma_{w}^{2} \mathds{1}(s=s').$ Essentially this assumption means that the error process can be represented as the sum between an error component with smooth covariance function and an IID white noise component. Let $\Xi(s,s')=\Sigma (s, s') + \Gamma(s,s')$, which is too a proper covariance function, and denote by $\{\phi_{k}(\cdot), \lambda_{k}\}_{k}$ its eigen-components. Using this basis we represent  $X_{ij}(\cdot)$ by $X_{ij}(\cdot) = \sum_{k = 1}^{K} \xi_{ijk} \phi_{k}(\cdot) $ where the basis coefficients are $\xi_{ijk} = \int_{\mathcal{S}} X_{ij}(s) \phi_{k}(s) ds.$ Let $K$ be a finite truncation; this approach implicitly assumes that the $K$ main eigenbasis functions are the most informative to predict the response. The assumption, that the components with the largest variation are most predictive of the dependent variable, is rooted in the principal component regression literature \citep{mardia1980multivariate}, and has been commonly employed in functional regression \citep{reiss2007functional,crainiceanu2009generalized, febrero2017functional}. Nevertheless it may be viewed as a strong limitation, and
future research is needed to investigate alternative approaches to select the orthogonal basis in a manner that appropriately accounts for the correlation between the functional predictor and response.

\subsection{Statistical modeling of the non-linear effects}
\label{sec:estiamtion_dynamic}

The univariate functions in model (\ref{eqn:integral:finite_summation_expansion1}), $\alpha(t)$ and $\beta_{1}(t) \ldots \beta_{K}(t)$ are unknown smooth functions. Assume for now that $X_{ij}(\cdot)$'s and furthermore $\xi_{ijk}$'s are known. The implied approximating model is known in the statistical literature as a time-varying coefficient model \citep{hastie1993varying, hoover1998nonparametric}. We briefly review it next and focus on how we ensure that the $K$ regression coefficients have the same smoothness. 

We use basis expansions - the truncated polynomial splines, B-splines or Fourier basis etc - to model the smooth parameter functions in (\ref{eqn:integral:finite_summation_expansion1}).
Let $\{B_{0l}(t)\}_l$'s and $\{B_{kl}(t)\}_l$'s be such bases and let $\alpha(t) = \sum_{l=1}^{L_{\kappa 0}} \beta_{0l} B_{0l}(t)$ and $\beta_k(t) = \sum_{l=1}^{L_{\kappa k}} \beta_{kl} B_{kl}(t)$.  For simplicity of exposition, we illustrate on truncated polynomial spline basis and take the bases to be the same, that is $L_{\kappa 1}=\ldots =L_{\kappa K}$ and $B_{0l}(\cdot)=B_{kl}(\cdot)=B_l(\cdot)$ for all $l\geq 1$. Let $\alpha(t) = \beta_{00} + \beta_{01} t + \ldots +\beta_{0p}t^p + \sum_{l=1}^{L} u_{0l} (t-\kappa_{l})^p_+$ and  $\beta_k(t) = \beta_{k0} + \beta_{k1} t + \ldots +\beta_{kp}t^p + \sum_{l=1}^{L} u_{kl} (t-\kappa_{l})^p_+$ where $\beta_{k0},\ldots, \beta_{kp}$'s are unknown fixed parameters, $u_{kl}$'s are independent random variables. 
Here $\kappa_{1}, \ldots, \kappa_{L}$ are knots in $\mathcal{T}$ and $(x)^p_+=\max(0, x^p)$. Typically,  
the coefficients of the non-polynomial functions are assumed to vary according to $u_{kl}\sim N(0, \sigma_k^2)$ for $k=0,1,\ldots, K$, where the variance $\sigma^2_k$ controls the smoothing of the unknown functions, $\alpha(\cdot)$ or $\beta(\cdot)$'s; see \cite{ruppert2003semiparametric}. As argued in Section \ref{sec:method}, estimating the unknown function using the  penalized criterion (\ref{eqn:penalized_criterion_general}) entails assuming same smoothness for the functions $\beta_k(\cdot)$'s for $k=1, \ldots, K$ or equivalently $\sigma_k^2=\sigma^2 $ for $k\geq 1$, $\sigma^2$ denotes their common value. 

This yields the following mixed effects representation of $g(\mu_{ij})=V_{ij} \beta + Z_{ij,0}u_0 +  \xi_{ij1}Z_{ij,1}u_1 +\ldots + \xi_{ijK}Z_{ij,K}u_K$, 
where $ V_{ij}=[1, t_{ij}, \ldots, t_{ij}^p , \xi_{ij1}, t_{ij} \xi_{ij1}, \ldots,   t^p_{ij}\xi_{ijK}]$ is a $(p+1)(K+1)$-dimensional row vector, $\beta=(\beta_{00}, \beta_{01}, \ldots, \beta_{0p},\beta_{10}, \beta_{11}, \ldots, \beta_{Kp})^T$ is the full vector of fixed effects. Also let $Z_{ij,k}$ is the $L$-dimensional row vector of $(t_{ij}-\kappa_{l})_+$'s and $u_{k}=(u_{k1}, \ldots, u_{kL})^T$ be the vector of random effects. Then $u_0 \sim N(0_L, \sigma^2_0 I_L)$ and $u_k \sim N(0_L, \sigma^2 I_L)$ for $k=1, \ldots, K$, where $I_L$ is the $L\times L$ identity matrix.  By an abuse of notation let $\xi_{ij0}=1$ for all $i$ and $j$; then (\ref{eqn:integral:finite_summation_expansion1}) becomes
\begin{equation}
\label{eqn:integral:finite_summation_expansion}
g(\mu_{ij})= V_{ij} \beta + \sum_{k=0}^K\xi_{ijk}Z_{ij,k}u_k  +Z_{b,ij} b_i.
\end{equation}

\section{Estimation}\label{sec:estimation}

In this section we detail the estimation of the model components, which is separated into covariates-related components, such as $\phi_k(\cdot)$ and $\xi_{ik}(\cdot)$, and response-related components, such as $\beta$, $u$, and $b$.  Prediction of the response trajectory $Y_i(\cdot)$ is detailed in Section \ref{sec:prediction}.

\subsection{Estimation of the covariates-related components}\label{ssec:phiscsi}

Modeling the functional covariates is done as in \cite{park2015longitudinal}. We briefly describe it here for completeness. We model the mean of $W_{ij}(s)$ at time $t_{ij}$ as a bivariate smooth function, and fit a bivariate smoother to estimate it, under a working independence assumption \citep{wood2006generalized}. We then demean the observed functional predictor; denote the demeaned data by $\widetilde W_{ij} (\cdot) $. Next $\widetilde  W_{ij}(\cdot)$'s is used to estimate the marginal covariance function $\Xi(s,s')=\Sigma( s, s') + \Gamma(s,s').$ The pooled sample covariance, defined as $\widetilde\Xi(s_{r},s_{r'}) =  \sum_{i=1}^{I} \sum_{j=1}^{n_{i}} \widetilde  W_{ijr} \widetilde  W_{ijr'} / (\sum_{i=1}^{I} n_{i})$, is a method of moments estimator of $\Sigma( s_r, s_{r'}) + \Gamma(s'_r,s_{r}) + \sigma_{w}^{2} \mathds{1}(r=r')$. This estimator is not smooth and may be viewed as a raw estimator of $\Xi(s'_r,s_{r}).$ The diagonal terms of $\widetilde\Xi(s_{r},s_{r'})$ are possibly inflated. One can ignore the diagonal terms and smooth the off-diagonal terms using a bivariate smoother \citep{yao2005functional, xiao2013fast}; denote the covariance estimator by $\widehat\Xi(s,s')$. We use \citep{xiao2013fast} for our numerical investigation. We estimate the eigen-components of $\Xi(s,s')$ by the eigen-components of $\widehat\Xi(s,s')$, $\{\widehat \phi_{k}(\cdot), \widehat \lambda_{k}\}_{k}$, where $\int_{\mathcal{S}} \widehat \phi_{k} (s) \widehat \phi_{k'} (s) ds = 1 $ if $ k = k' $ and 0 otherwise, and $\widehat\lambda_{1} \geq \widehat\lambda_{2} \cdots \geq 0$. Let $K$ be so that the first $K$ pairs provide a low-rank approximation of  $\widehat\Xi(s,s')$: $\widehat\Xi(s,s') \approx \sum_{k=1}^{K} \widehat \lambda_{k} \widehat \phi_{k}(s) \widehat \phi_{k}(s')$. %Note that the truncation value $K$ is chosen using the procedure described in \cite{staicu2010fast}. 
Using numerical integration, the time-varying loadings $\widetilde \xi_{{W}, ijk}$'s are estimated as $\widetilde \xi_{{W}, ijk}= \int_{\mathcal{S}} \widetilde W_{ij}(s) \widehat \phi_{k}(s) ds$ for $k = 1, \cdots, K.$ Nevertheless these quantities are noisy estimates of $\xi_{ijk}$ and regressing the response directly onto them would lead to increased bias in the estimates. In addition, they correspond to the times $t_{ij}$ solely and would not be suitable to be used when predicting the response trajectory $Y_i(t)$ for any time point $t$ is of interest. We propose to model $\widetilde \xi_{{W}, ijk}$ in a way that explicitly recognizes the dependence on the time $t_{ij}$.

Consider the working model $\widetilde \xi_{{W}, ijk} =  \xi_{ik}(t_{ij}) + \epsilon_{W,ijk}$, where $\xi_{ik}(\cdot)$ is a random curve with zero mean and covariance function $G_k(\cdot, \cdot)$ such that $G_k(t_{ij}, t_{ij'})=\textrm{cov}\{\xi_{ijk}, \xi_{ij'k}\}$, and $ \epsilon_{W,ijk}$ is white noise with zero mean and finite variance $\sigma^2_{W,k}$. Here $\xi_{ijk} = \xi_{ik}(t_{ij}).$  Recovering the trajectories $\xi_{ik}(\cdot)$ requires modeling and estimation of their covariance function. In this regard, we use the pseudo data  $\{(\widetilde \xi_{{W}, ijk}, t_{ij}): j=1, \ldots, n_i\}_{i}^{I}$, separately for each $k$. One simple approach is to assume a parametric covariance model, such as exponential, or Mat\'ern, or random effects based models; see \cite{park2015longitudinal} for more discussion. Standard methods in longitudinal data analysis can be used to estimate the covariance model. Here we consider a flexible nonparametric covariance model as it is common in functional data analysis and employ common techniques in sparse functional principal components analysis \citep{yao2005functional} to estimate it; this approach was described by \cite{park2015longitudinal}. The spectral decomposition of $G_{k}(\cdot,\cdot)$ is $G_{k}(t,t') = \sum_{l \geq 1} \eta_{kl} \psi_{kl}(t) \psi_{kl}(t')$, where $\{\eta_{kl}, \psi_{kl}(\cdot)\}$ is the pair of eigenvalues and eigenfunctions for $\eta_{k1} \geq \eta_{k2} \geq \cdots > 0,$ and $\{\psi_{kl}(\cdot)\}_{l \geq 1}$ are mutually orthogonal, and have unit norm in $L^{2}(\mathcal {T}).$ Using the truncated Karhunen-Lo\`eve (KL) expansion $\xi_{ik}(t) = \sum_{l \geq 1} \zeta_{ikl} \psi_{kl}(t)$, where $ \zeta_{ikl} = \int_{\mathcal{T}} \xi_{ik}(t) \psi_{kl}(t)dt$ is random, with zero-mean and variance equal to $\eta_{kl}$.

Let $\widehat G_k(\cdot, \cdot)$ be a covariance estimator of $G_k(\cdot, \cdot)$ obtained as \cite{yao2005functional, crainiceanu2009generalized}. The spectral decomposition of $\widehat G_{k}(t,t')$, $\widehat G_{k}(t,t') \approx \sum_{l=1}^{L_{k}} \widehat \eta_{kl} \widehat \psi_{kl}(t) \widehat \psi_{kl}(t')$ yields orthogonal functions that have unit norm, $\widehat \psi_{kl}(\cdot)$'s and non-negative eigenvalues, $\widehat \eta_{kl}$'s. Here $L_{k}$'s are truncation values that are chosen in similar style as $K$. The time-varying loadings are estimated using the truncated Karhunen-Lo\`eve expansion $\widehat \xi_{ik}(t)  =  \sum_{l=1}^{L_{k}}  \widehat \zeta_{ikl} \widehat \psi_{kl}(t) $; the scores $\widehat \zeta_{ikl} $ are obtained via conditional expectation $E[\zeta_{ikl}|\widetilde \xi_{W,i1k}, \ldots, \widetilde \xi_{W,in_ik}]$ in the associated mixed effects model $\widetilde \xi_{W,ijk} =  \sum_{l=1}^{L_{k}}    \zeta_{ikl} \widehat \psi_{kl}(t) + e_{W, ijk}$ and using a working Gaussian response assumption.

\subsection{Estimation of the response-related components}
\label{ssec:betasu}

The estimation of the model parameters $\alpha(\cdot)$ and $\beta_k(\cdot)$'s for $k=1, \ldots, K$ in (\ref{eqn:integral:finite_summation_expansion1}), using the basis representation described in the Section \ref{sec:estiamtion_dynamic}, entails estimation of $\beta$'s and $u_k$'s in (\ref{eqn:integral:finite_summation_expansion}), where $\xi_{ijk}$ are replaced by $\widehat \xi_{ ik}(t_{ij})$.  For exposition simplicity denote by  
${\widetilde Z_{ij,k} = \widehat \xi_{ ik}(t_{ij}) Z_{ij,k}}$ for all $i,j$ and $k$ and let $\widetilde V_{ij}$ be the vector $V_{ij}$ with {$\widehat \xi_{ ik}(t_{ij})$}'s used in place of $\xi_{ijk}$'s.

It follows that $g(\mu_{ij}) =\widetilde V_{ij} \beta + \sum_{k=0}^K\widetilde Z_{ij,k}u_k+  Z_{b,ij}  b_{i}$; remark that the subject specific effects $b_i$'s account for the dependence across repeated observations within the same subject. Then the model for $Y$ can be approximated by the following generalized mixed effects model
\begin{eqnarray}
\label{eqn:estimation_parameters}
& &Y  \sim EF(\mu, \eta) \nonumber \\
& & \mu_g= \widetilde V\beta + \widetilde Z u +  Z_{b}  b, \nonumber \\
& & u \sim N \left(0_{L+LK},  \textrm{diag}\{ \sigma_0^2,  \sigma^2, \ldots,  \sigma^2\}\otimes I_{L} \right ) \text{ and }
b\sim N \left(0_{Iq}, I_I \otimes D\right) \\
&& u \text{ and } b \text{ are mutually independent}   \nonumber 
\end{eqnarray}
where $\mu_g$ is obtained by columnwise stacking $(g(\mu_{i1}), \ldots, g(\mu_{in_i}))^T$, $\widetilde V$ is obtained by stacking columnwise $\widetilde V_{ij}$ first over $j$ and then over $i$. Here $u=(u_0^T|u_1^T|\ldots|u_K^T)^T$ and $b=(b_1^T |\ldots|b_I )^T$, $\widetilde Z = (\widetilde Z_0| \widetilde Z_1| \ldots | \widetilde Z_K)$ with $\widetilde Z_k$ obtained like $\widetilde V$ by stacking columnwise $Z_{ij,k}$ over $j$ and $i $ and $Z_b = \textrm{diag} \{Z_{b,1}, \ldots, Z_{b,I} \}$ and $Z_{b,i}$ is obtained by stacking columnwise $Z_{b, ij}$ over $j=1, \ldots, n_i$. Once the model is represented in this form, parameter estimation and quantification of their estimation uncertainty follows easily.

For given values of the covariance parameters, $\sigma^2_0$, $\sigma^2$, and $D$, the estimates of $\beta$, $u_k$'s and $b_i$'s are the same as the minimizers of the following penalized criterion:

\begin{eqnarray} 
\label{eqn:penalized_criterion}
pl( \beta, u_0, \ldots, u_K, b, \eta ) =   -2\sum_{i=1}^I\sum_{j=1}^{n_i} \ell_{ij}( \beta, u_0, \ldots, u_K, b, \eta)  + b^T (I_I \otimes D^{-1}) b + 
+ \lambda_0\|u_0\|^2 +\lambda \sum_{k=1}^K\|u_k\|^2,
\end{eqnarray}

where $\ell_{ij}( \beta, u_0, \ldots, u_K, b, \eta) $ is the log-likelihood function corresponding to the assumed conditional model for $Y_{ij}$, $\lambda_0=1/\sigma^2_0$, and $\lambda=1/\sigma^2$. Modeling the smoothness of the unknown functions explicitly, as the inverse of a variance component, allows us to clearly describe the corresponding generalized mixed effects model (\ref{eqn:estimation_parameters}). The criterion (\ref{eqn:penalized_criterion}) can be easily modified to account for other bases and associated penalties: other choices of bases will modify the term $\widetilde V_{ij} \beta +\sum_{k=0}^K\widetilde Z_{ij,k} u_k$ that appears in the expression of $\ell_{ij}( \beta, u_0, \ldots, u_K, b, \eta)  $, while the associated penalties will modify the term $\lambda_0\|u_0\|^2 +\lambda\sum_{k=1}^K \|u_k\|^2$. The smoothing parameters $\lambda_0$ and $\lambda$ will continue to have the same interpretation; see \cite{wood2006generalized} and \cite{ivanescu2015penalized}.  

To estimate the variance parameters, a Bayesian perspective where the parameters are estimated using the log of a corresponding marginal likelihood, is more appealing. The ideas are described in \cite{wood2011fast} and \cite{wood2016smoothing} and rely on using the Laplace approximation to calculate the desired marginal log-likelihood. When the responses are Gaussian, the implied marginal log-likelihood corresponds to the restricted maximum likelihood (REML). For both our simulation study and data application we used REML to select the smoothing parameters. 

Using the parameter estimates from (\ref{eqn:penalized_criterion}) we obtain estimates of the intercept function $\widehat \alpha(t) = \widehat \beta_{00} + \ldots + \widehat \beta_{0p}t^p + \sum_{l=1}^{L} \widehat u_{0l} (t-\kappa_{l})_+$ and of the regression bivariate function
\begin{eqnarray}
\widehat \gamma(s,t) =\sum_{k=1}^K \widehat \phi_k(s) \widehat \beta_k(t),
 \label{eqn:param_fns_estimates}
\end{eqnarray}
where 
$\widehat \beta_k(t) = \widehat \beta_{k0} + \ldots + \widehat \beta_{kp}t^p + \sum_{l=1}^{L}\widehat u_{kl} (t-\kappa_{l})_+$ for $k=1, \ldots, K$. Then, conditional on the esitmates from the pre-processing of the functional covariates, uncertainty quantification for $\widehat \alpha(\cdot)$ and $\widehat \gamma(\cdot, \cdot)$ is readily available. However, inference for $\gamma(s,t)$ is meaningless, due to the lack of parameter identifiability in the model (\ref{eqn:LFR_model}). 

In this paper we focus on response prediction and quantify the prediction uncertainty. There are many sources of uncertainty: the estimation of the basis functions $\phi_k(\cdot)$, of the basis functions coefficients $\xi_{ijk}$'s and of their covariance estimators, as well as the various truncations $K$ and $L_k$'s. It is not clear how to account for all these sources in estimating the uncertainty of prediction. In the next section we discuss prediction and its associated inference, conditional on all these quantities.

\section{Prediction and inference}\label{sec:prediction}

\subsection{Prediction of response trajectories}

One of our main aim is to predict response trajectories in two settings: for an existing data subject, $i$, who has been observed at few sparse time points $t_{ij}$'s, and for a new subject, $i^*$ whose only functional observations are available. In general, the response for an existing data subject at an observed time $t_{ij}$, $Y_{ij} $, can be predicted by directly substituting the estimates of the parameters and the predicted random effects, $\widehat b_i$ into equation (\ref{eqn:estimation_parameters}); the specifics depend on the form of the link function. Irrespective of whether prior response data has been observed for a subject, prediction of a subject's  response trajectory $Y_{i}(t)$ for all $t$ requires estimation of the subject's time-varying basis coefficients trajectories $\xi_{ik}(t)$; recall $\xi_{ik}(t_{ij}) = \xi_{ijk}$.

Let $\{W_i(\cdot, t_{ij}), j=1, \ldots, n_i\}$  be the noisy functional covariate for a subject already in the data, and denote by $Y_{i1}, \ldots, Y_{in_i}$ the associated responses; it is assumed that $Y_{ij}=Y_i(t_{ij})$, $W_i(\cdot,t)$ is a noisy measurement of $X_i(\cdot, t)$, and that $Y_i(t)$ relates to $X_i(\cdot, t)$ through the model $Y_i(t) \sim EF (\mu_i(t), \eta)$, where $g\{\mu_i(t) \}= \alpha(t) +\int \gamma(s,t) X_i(s,t) ds + Z_{b,it} b_i +\epsilon_{it}$. Here $Z_{b,it}$ is the random design vector corresponding to a generic time $t$ and $\epsilon_{it}$, denoted by an abuse of notation, is a white noise process with zero mean and variance $\sigma^2$. For example, in the case of a random effects that involves a random intercept and slope, $b  =(b_{0}, b_{1})^T$, we have $Z_{b,it} b_i= b_{i0}  +  b_{i1} t$.

The subject mean response trajectory for an existing data subject, $\mu_i(t)$, is predicted by:
\begin{eqnarray}
\label{prd_full}
\widehat \mu_{i}(t) =  g^{-1}\{\widehat \alpha(t) +\sum_{k=1}^K \widehat \xi_{ik}(t) \widehat \beta_k(t) + Z_{b,it} \widehat b_{i}  \},
\label{eqn:prediction}
\end{eqnarray}
where $g^{-1}$ is the inverse function of $g$. For Gaussian responses, $g(\mu)=\mu$, expression (\ref{eqn:prediction}) can be used to predict subject trajectories $\widehat Y_i(t) = \widehat \mu_{i}(t)$. We conjecture that this predicted subject trajectory is a consistent estimator of $\widetilde Y_{i}(t)=E[Y_i(t)|W_i] $. In the case considered here, $\widetilde Y_{i}(t) =\alpha(t) +\sum_{k\geq1} \widetilde \xi_{ik}(t) \beta_k(t) +  Z_{b,it}  b_{i}$, where $\widetilde \xi_{ik}(t) = \sum_{l\geq 1} \psi_{kl} (t) \widetilde \zeta_{ikl} $ and $\widetilde \zeta_{ikl} = E[\zeta_{ikl}| \xi_{ijk}: j=1, \ldots, n_i]$. When the responses are non-Gaussian, prediction of the subject specific trajectories is not always clearly defined. For Bernoulli responses a common approach is to predict $\widehat Y_{i}(t)=1$ if $\widehat \mu_{i}(t) \geq 0.5$ and predict $\widehat Y_{i}(t)=0$, if $\widehat \mu_{i}(t) < 0.5$.

For a new subject $i^*$, conditional on the functional covariates $\{W_{i*j}(\cdot) = W_{i^*}(\cdot, t_{i^*j}):j\}$, the mean response trajectory is predicted as
$\widehat \mu_{i^*}(t) = g^{-1}\{ \widehat \alpha(t) +\sum_{k=1}^K \widehat \xi_{i^*k}(t) \widehat \beta_k(t)\}$, where the time-varying trajectories $\widehat \xi_{i^*k}(t)$ are obtained as presented in Section \ref{ssec:phiscsi} using the noisy pseudo-data $\widetilde \xi_{W,i^*jk} = \int \widetilde W_{i^*j}(s) \widehat \phi_k(s)ds$ and assuming a working model $\widetilde  \xi_{W,i^*jk} = \sum_{l=1}^{L_k} \widehat \psi_{kl}(t) \zeta_{i^*kl} + \epsilon_{i^*jk}$, where $\zeta_{i^*kl} \sim N(0, \widehat \eta_{kl})$ and $\epsilon_{i^*jk}\sim N(0, \widehat \sigma^2_{\epsilon_k})$. The response trajectory does not involve subject-specific effects $b_{i^*}$, as their estimation requires availability of response data at repeated times.

\subsection{Asymptotic prediction bands}\label{ssec:pred_bands}

For Gaussian responses, we can construct asymptotic prediction bands for the individual response trajectory, conditional on the underlying predictor function. The prediction bands do not account for the variability associated with the estimation of the basis functions $\{ \widehat \phi_k(\cdot): k= 1, \ldots, K\}$, the truncation $K$, $\{ \widehat \psi_{kl}(\cdot): l\geq 1\}$, $\{\widehat \eta_{kl} : l\geq 1\}$,  and $L_k$ for $k=1, \ldots, K$. %We start with prediction of the response trajectory for a subject whose response was observed at multiple times and then discuss prediction for a new subject with only functional covariates available. 

Let $\{W_{ij}(\cdot), t_{ij}: j=1, \ldots, n_i\}$ be the observed functional covariates for a subject $i$. The uncertainty in prediction is measured by the prediction error \citep{ruppert2003semiparametric} $\{ \widehat Y_i(t)-Y_i(t)\}$; for a new subject, or in the case of an existing subject for $t\nin \{ t_{i1}, \ldots,t_{in_i}\}$ we have
\begin{eqnarray}
{\textrm{Var}}\{ \widehat Y_i(t)-Y_i(t)\} =  {\textrm{Var}}\{ \widehat Y_i(t)\}  + {\textrm{Var}}\{ \epsilon_{it} \}.
\label{eqn:iterative_var}
\end{eqnarray}
The variance of $\{ \epsilon_{it} \}$ is estimated using REML along with the other variance parameters; see Section \ref{ssec:betasu}. The variance of $\{ \widehat Y_i(t)\} $ is estimated with standard approaches in longitudinal data analysis \citep{ruppert2003semiparametric, wood2006generalized, wood2006confidence}; the estimation is implemented in various computer packages and we discuss it in the Supplementary Material, Section C. Thus a $100(1-\alpha)\%$ pointwise prediction interval for $Y_i(t)$ is $\widehat Y_i(t) \pm z_{\alpha/2} \widehat SE\{ \widehat Y_i(t)-Y_i(t)\} $, where $z_{\alpha/2}$ is the $\alpha/2$ upper quantile of the standard normal distribution. Here $\widehat SE\{ \widehat Y_i(t)-Y_i(t)\} $ is the estimated standard error of $\{\widehat Y_i(t)-Y_i(t)\} $ and is calculated as the square root of the estimated variance of $\{ \widehat Y_i(t)-Y_i(t)\}$. The terms  $\widehat Y_i(t)$  and $\widehat SE\{ \widehat Y_i(t)-Y_i(t)\} $ have different expression according to whether the subject is an existing data subject, or is a new subject, along similar lines as detailed in the previous subsection. In particular, in the case of new subject, they do not include estimates of the random subject effects $b_i$ and their estimation variability. In Section \ref{ssec:simsetting:pred} we assess the performance of the response trajectory as well as that of the proposed pointwise prediction intervals for both existing data subjects and new subjects.

% These confidence bands are conditional on W, and they are constructed to ensure a coverage of at least 1-alpha. Thus the unconditional probab is also at least 1-alpha. That's bc P(X) = E[1(X)]=EE(1(X)|W)

\section{Simulation} \label{sec:simulation}
\subsection{Description of the settings} \label{ssec:simsetting}

We use Monte Carlo simulations to assess the numerical performance of the proposed method (LDFR) and compare it with two other competing approaches: LPFR \citep{goldsmith2012longitudinal} and LPEER  \citep{kundu2016longitudinal}. 
The data $[ t_{ij}, Y_{ij}, \{ (W_{ijr}, s_{r}): r = 1, \cdots R\}: j = 1, \cdots, n_{i} ]^{I}_{i=1}$ are generated according to the following scenarios:
\begin{itemize}
	\item[($A$)] $X_{i}(s,t) = \tau(s,t) + \sqrt{2} \zeta_{i11} \cos(2\pi t) \cos(2 \pi s) + \sqrt{2} \zeta_{i12} \sin(2\pi t) \cos(2 \pi s) +  
	\sqrt{2} \zeta_{i21} \cos(4\pi t) \sin(2 \pi s) + \sqrt{2}  \zeta_{i22} \sin(4\pi t) \sin(2 \pi s)$, 
\end{itemize}	
 where $\tau(s,t) = 1+2s+3t+4st.$ Moreover, $\zeta_{i11}$, $\zeta_{i12}$, $\zeta_{i21}$, and $\zeta_{i22}$ are assumed to be mutually independent and identically distributed (IID) such as $\pN(0, 3.5)$, $\pN(0, 2)$, $\pN(0, 3)$,  and $\pN(0, 1.5)$ respectively. Let $W_{ijr} = W_{i}(s_{r}, t_{ij}) = X_{i}(s_{r}, t_{ij}) + \epsilon^{w}_{ij}(s_{r})$. We define the error term as $\epsilon^{w}_{ij}(s_{r}) =  \sqrt{2}  \cos(2 \pi s_{r}) \epsilon_{1,ij} + \sqrt{2}  \sin(2 \pi s_{r}) \epsilon_{2,ij} + \epsilon_{3,ij}(s_{r}).$ Here, $\epsilon_{1,ij}, \epsilon_{2,ij},$ and $\epsilon_{3,ij}(s_{r})$ are IID such as $\pN (0, \sigma^{2}_{\epsilon_{1}})$, $ \pN (0, \sigma^{2}_{\epsilon_{2}})$, and $\pN (0, \sigma^{2}_{\epsilon_{3}})$; where, $\sigma^{2}_{\epsilon_{1}} = 0.3$, $\sigma^{2}_{\epsilon_{2}} = 0.7$, and $\sigma^{2}_{\epsilon_{3}}$ are calculated using signal-to-noise-ratio (SNR) which is defined as $$ SNR = \frac{ \int_{\mathcal{T}} \int_{\mathcal{S}} var \{W_{i}(s,t)\} ds dt}{\sigma^{2}_{\epsilon_{1}} + \sigma^{2}_{\epsilon_{2}} + \sigma^{2}_{\epsilon_{3}} } - 1.$$    
\begin{itemize}
	\item[($B1$)]  {\itshape Large noise variance}: $\sigma^{2}_{\epsilon_{3}} = 9$ (i.e. $SNR = 0.5$).
	\item[($B2$)]  {\itshape Small noise variance}: $\sigma^{2}_{\epsilon_{3}} = 1$ (i.e. $SNR = 2.5$).
\end{itemize}
We consider a dense design for $s$: $\{s_1, \ldots, s_R\}$ is taken as a grid of $101$ equidistant points in $[0,1]$ and two sampling designs for $t_{ij}$'s depending on number of repeated measurements $n_i$: 
\begin{enumerate}
	\item[($C1$)] {\itshape Sparse design} when the number of repeated responses per subject is: $n_{i} \in \{6,\cdots, 10\},$
	\item[($C2$)] { \itshape Moderately sparse design} when $n_{i} \in \{16, \cdots, 20\}.$
	%\item[($C3$)] Balanced design: $n_{i} = n;$ where $n$ = 16 or $n$ = 31. 
\end{enumerate}  
In each case $\{t_{i1}, \cdots, t_{in_{i}} \}$ are randomly chosen from a set of 41 equidistant points in $ [0,1]$. \\

\noindent  We generate $Y_{ij}$ in the exponential family as follows: 
\begin{itemize}
	\item[($D1$)] Gaussian responses with mean $\mu_{ij}=\mu_i(t_{ij})$ defined by %with identity link function, i.e. $g(\mu_{ij}) = \mu_{ij}$
	 $\mu_{i}(t) =\alpha(t)+\int X_i(s, t) \gamma^{\delta}(s,t)  ds  $, where $\alpha(t) = 7 \sin(3 \pi t)$ and random deviation $\epsilon_{ij}$. Consider three dependence structures:
\begin{itemize}	
	\item[(i)] {\itshape Independent covariance structure:} $\varepsilon_{ij}$ is distributed as IID $\pN(0, 2).$
	
	\item[(ii)] {\itshape Compound symmetric (CS) structure:} $\varepsilon_{ij} = b_{i0} + e_{ij}$, where, $b_{i0}$ and $e_{ij}$ are distributed as IID $\pN(0, 1)$ and $\pN(0, 0.5)$ respectively, and are mutually independent. 
	
	\item[(iii)] {\itshape Random effect model (REM):} $\varepsilon_{ij} = b_{i0} + b_{i1}t_{ij} + e_{ij}$, where, $b_{i0}$ and $b_{i1}$ are distributed as IID $\pN (0, 1)$ and $\pN (0, 0.5)$ respectively with $cov(b_{i0}, b_{i1}) = 0.1$. Also $e_{ij}$ is distributed as IID $\pN(0, 0.3)$, and is independent from both $b_{i0}$ and $b_{i1}.$ 
	
\end{itemize}
\end{itemize}

\begin{itemize}	
	\item[($D2$)] Binary-valued responses. We consider $P(Y_{ij}=1) = exp( \omega_{ij})  / \{ 1 + exp( \omega_{ij}) \},$ where $\omega_{ij} = \alpha(t_{ij}) + \int_{\mathcal{S}} X_{i}(s, t_{ij})\gamma^{\delta}(s,t_{ij})ds + b_{i0} + b_{i1}t_{ij}$, and the subject-specific random effects $b_{i0}, b_{i1}$ are used to model the dependence across the repeated measurements and are generated as in (iii) above. 
	The choice for $\alpha(\cdot)$ described in $(D1)$ varies with time; while a time-varying intercept presents no issues for our proposed method, such choice does not seem to be accommodated by LPFR. Thus, for binary-valued responses, in order to compare the results of our method to LPFR, we consider $\alpha(\cdot) = 2$.	
\end{itemize}

We consider the functional coefficients as below:
\begin{itemize}	
	\item[($E1$)] Mixture of trigonometric and exponential functions of $t$: $\gamma^{\delta}(s,t) = \sqrt{2} \exp (-\delta t) cos(2 \pi s) + \sqrt {2} \delta t \sin ( \delta t) sin(2 \pi s)$. 
	\item[($E2$)] Polynomial function of $t$: $\gamma^{\delta}(s,t) = \sqrt{2} (1 + \delta t ) cos(2 \pi s) + \sqrt {2} (1 - \delta t + \delta t^2 )  sin(2 \pi s)$.
\end{itemize}	
In both cases the parameter $\delta$ controls the departure from a time-invariant effect. For example in ($E1$) when $\delta = 0$ we have that $\gamma^{\delta}(s,t) = \sqrt{2} cos(2 \pi s),$ while when $\delta \neq 0,$ $\gamma^{\delta}(s,t)$ varies with time $t$. We investigate the cases $\delta = \{ 0, 1, 2, 5, 10\}.$ The various settings amount to a signal to noise ratio of the response varying between 0.1 (B1, $\delta=1$, and any $D1i-D1iii$)  to 3.5 (B2, $\delta=5$ and any $D1i-D1iii$) as defined in Section A1 of the Supplementary Material.

We study the performance of our proposed method for varying sample sizes $I \in \{ 100, 200, 300 \}.$ The implementation of the methodology consists of two main steps. First pre-process the noisy longitudinal functional covariates using \cite{park2015longitudinal}; implemented in the function {\tt fpca.lfda} in the {\tt R} package {\tt refund} (\cite{refund}). For transparency we describe it briefly:  (1)  Estimate the bivariate mean function $\widehat \tau(\cdot,\cdot)$ using the fast bivariate smoother based on tensor product of cubic splines with 35 knots in each direction and second order difference penalty \citep{xiao2013fast}, and by selecting the smoothing parameter using GCV. (2) Demean the observed functional predictors and estimate the smooth marginal covariance using the sandwich bivariate smoother. (3) Perform eigenanalysis of the estimated smooth covariance and obtain the pairs of eigenvalues and eigenfunctions $\{ \widehat \lambda_{k}, \widehat \phi_{k}(\cdot)\}^{K}_{k}$; here $K$ is chosen using 95\% PVE value. (4) Estimate the time-varying loadings as $\widetilde \xi_{ijk} = \int_{\mathcal S} \{W(s, t_{ij}) - \widehat \tau(s,t_{ij})\} \widehat \phi_{k}(s) ds$ using numerical integration. (5) For each $k$, consider $\{\widetilde \xi_{ijk}, t_{ij}:j=1, \ldots, n_i\}_i$ and assume a nonparametric covariance structure for the dependence of $\widetilde \xi_{ijk}$ across $j$ as described in Section \ref{ssec:phiscsi}. Furthermore, estimate the eigen-components $\{\widehat { \eta}_{kl}, \widehat { \psi}(\cdot)_{kl} \}^{L_{k}}_{l=1}$ where $\eta_{kl}$ are the non-decreasing non-negative eigenvalues; $L_{k}$ is chosen by 95\% PVE value. Predict the scores $\zeta_{ikl}$ using the associated truncated mixed effects model discussed in Section \ref{ssec:phiscsi} and calculate the estimated time-varying loadings for any time $t \in \mathcal{T}$; i.e. $\widehat \xi_{ik}(t) = \sum_{l=1}^{L_{k}} \widehat \zeta_{ikl} \widehat \psi_{kl}(t)$. 
The second step uses $\widehat \xi_{ijk} = \widehat \xi_{ik}(t_{ij})$ in  (\ref{eqn:integral:finite_summation_expansion}) and fits the approximated generalized mixed model with penalties and assuming independent random effects; identity link for Gaussian responses and logit link for binary responses. The function {\tt lme} in the {\tt R} package {\tt nlme} is used at this step.

For LPFR the time-invariant regression coefficient $\gamma(s)$ is modeled using the truncated linear basis with $30$ functions (default choice) and the smoothing parameter is estimated by REML; the model is fitted using the function {\tt lpfr} in the {\tt R} package {\tt refund}. For LPEER, the time-varying coefficient $\gamma(s, t)$ is modeled using a polynomial basis in time $t$, with coefficients that are smooth functions in $s$ and which are estimated using a penalized criterion with a second-order difference penalty; the degree of the polynomial basis is selected using the Akaike information criterion (AIC) (\cite{akaike1974new}) and the smoothing parameters of the smooth terms are selected using REML. The model is fitted using the function {\tt lpeer} in the {\tt R} package {\tt refund}.

\subsection{Evaluation criteria}
 
To assess the prediction performance of the method, we divide each simulated dataset into a training and test set. Both sets contain information for the $I$ subjects; recall $I$ is the total number of subjects. The test set is formed as follows: for each subject $i$ in the dataset,  we randomly select five instances without replacement from the available $n_i$ instances, say $\{t_{ij_1}, t_{ij_2}, t_{ij_3}, t_{ij_4},t_{ij_5}\}$, and include the corresponding information $[ t_{ij}, Y_{ij}, \{(W_{ijr}, s_r): r=1, \ldots R\}, j=j_1,j_2, j_3, j_4, j_5]$ in the test data. The remaining observations for each subject are included in the training set. We fit the model using the training data; then we predict both the responses in the training set (IN) and the responses in the test set (OUT) using the estimates obtained from the fit on training data. To evaluate the performance of the competing models for normal responses, we compute the root-mean-prediction-error for the training set (IN$_{PE}$) and for the test set (OUT$_{PE}$); i.e. IN$_{PE} =  \sqrt{ \sum_{i=1}^{I} \{ \sum_{j\nin \{ j_1,\ldots, j_5\}} (Y_{ij} - \widehat Y_{ij} )^{2} / (n_{i} - 5) \} / I}$ and OUT$_{PE} = \sqrt{ \sum_{i=1}^{I} \{ \sum_{j\in \{ j_1,\ldots, j_5\}} (Y_{ij} - \widehat Y_{ij} )^{2} / 5 \} / I}$. For binary-valued responses, we assess the numerical performance in estimating the linear predictor trajectory $g(\cdot)$, and with respect to sensitivity or true positive rate (TPR); where TPR is defined as the proportion of successes ($\widehat Y_{ij}=1$) that are correctly identified. 

The prediction of the entire trajectory is assessed using the root mean prediction error,
$RMPE_{trj}$ of $\widehat Y_i(\cdot)$ which is defined as $RMPE_{trj}= \sqrt{1/I \sum_{i=1}^{I}[1/n \sum_{j=1}^{n} \{ Y_{i}(t_{j}) - \widehat Y_{i}(t_{j}) \}^{2}]},$ where $\{t_1, \ldots, t_n\}$ is an equally spaced grid of 41 points in $[0,1]$ and $Y_i(t_j)$ is obtained using the generating model. For this part, the model parameters are estimated using the entire data set, and not just the training data set. 

The accuracy of the pointwise prediction bands is evaluated in two cases. First, we assess the performance of the prediction bands for all the existing data subjects, that is subjects whose data are used to estimate the model parameters. Second, assess the performance for prediction bands of new subjects responses, whose functional predictor information is available solely. In the latter case, we construct a new set of $100$ subjects and for each set we generate data according to our model; the data for these subjects are not used in the estimation of the model parameters. 
In both cases the performance of the $100(1 - \alpha)\%$ pointwise prediction band, say $PB_t=(PB^{l}_t, PB^{u}_t)$ specified in terms of its endpoints and which is constructed as detailed in Section \ref{ssec:pred_bands}, is evaluated using the average pointwise coverage defined as
$
1 /I \sum_{i=1}^I \sum_{j=1}^n {1} \{ Y_{i}(t_j) \in PB_{t_j} \} /n
$,
where $1(x\in A)$ equals $1$ if $x\in A$ and $0$ otherwise. We also calculate the expected length of the constructed prediction bands as  $
1 /I \sum_{i=1}^I \sum_{j=1}^n (PB^{u}_{t_j} -PB^{l}_{t_j}) /n
$. 

The results are based on 1000 independent samples for each combination of the simulation settings. In our numerical investigation, we use Intel(R) Core(TM) i7-4770, 3.40 GHz processor with 8.0 GB RAM in 64-bit operating system.

\subsection{Prediction performance assessment and comparison}\label{ssec:simsetting:pred}

\emph{Prediction accuracy}. First, we consider Gaussian responses ($D1$) and compare our longitudinal dynamic functional regression method (LDFR) with LPFR. Table \ref{CS} displays the median of IN-and-OUT sample prediction errors for different $\delta$ values for 1000 simulations along with their respective interquartile ranges (IQR) in parenthesis. The results correspond to data generated using CS dependence structure ($D1ii$) and fitted by assuming a model with $CS$ type covariance structure. We observe that both types of prediction errors (IN and OUT) are similar for the two approaches when $\delta = 0$. However, as $|\delta| > 0$ the functional coefficient is time-dependent and the prediction results with the proposed method are superior relative to LPFR.  For example, when $\delta = 5$ our method yields improvement in prediction accuracy by more than 40\% over LPFR. Furthermore, the numerical study shows that the accuracy of our method increases with the number of repeated measurements per subject; this is expected as in this case, the estimation of the within subject covariance improves. In the Supplementary Material, Section A.2 we investigate mild misspecification of the dependence structure and observe similar findings.

Next, we compare the performance of LDFR with LPEER. Because of the heavy computational burden of the latter approach, we limit our investigation to 100 Monte Carlo samples per setting; see Table \ref{lpeer}. Here, we fit the competing model without assuming prior knowledge about the structure of the bivariate regression coefficient $\gamma(s,t)$. Table \ref{lpeer} illustrates the prediction performance when data are generated from $CS$ type covariance structure ($D1ii$). We fit the model by using a random subject intercept model (correct covariance model). When $|\delta|>0$ the departure of $\gamma(s,t) $ from a time-invariant coefficient is stronger. The numerical results show improved performance for our method as $|\delta|>0$. When $\delta = 1$,  LDFR and LPEER show similar prediction accuracy. However, LDFR is computationally over an order of magnitude faster than LPEER; see the third and sixth pairs of columns in Table \ref{lpeer}. Furthermore, for $\delta = 5$, LDFR outweighs LPEER in nearly all the cases considered. This is possibly due to the fact that LPEER, in its implementation %and, in the lack of prior information about the shape of the regression coefficient, 
models  $\gamma(s, \cdot)$ using a polynomial basis in $t$ and selects the number of basis functions from few choices; for the case ($E1$), a much richer polynomial basis in $t$ is needed to approximate $\gamma(s, \cdot)$, than the bases considered. In contrast, the proposed method does not rely on such assumption. As an anonymous reviewer suggested, we also compare the prediction accuracy of the two approaches when the true regression coefficient $\gamma(\cdot, \cdot)$ is a linear combination of polynomial functions in $t$, case ($E2$). The results are shown in Table 7 in the Supplementary Material, Section A2, and are consistent to the ones reported in Table \ref{lpeer}; the major difference is the improved computing time for LPEER, but still much higher compared to our method. Also, there seems to be some numerical stability issues with LPEER; in our simulation study we experienced convergence problems in few cases where the sample size is small.

We consider binary responses with logit link ($D2$) and evaluate the prediction accuracy of the proposed method with LPFR, which is the only existing alternative. We fit the model with subject-specific random intercept, while data are generated assuming both random intercept and slope (model misspecification). Table \ref{binarySNR5_larger_noise} shows the prediction error of the linear predictors $\widehat g(\mu_{ij})$ for both sparse and moderately sparse longitudinal designs, when the functional covariates are observed with large noise. The results are consistent to the previous ones. 
%: when $\delta = 0$, the numerical performance of both LDFR and LPFR is similar but when $\delta$ departs from zero, say for $\delta = 5$, the prediction accuracy of LDFR improves substantially. 
As the magnitude of $\delta$ increases the LDFR results also show an improved performance over LPFR with respect to the true positive rate. Additional simulation results for both Gaussian response and binary response cases are included in the Supplementary Material, Section A.3.

\begin{figure}[H]
	\caption{\small Gaussian responses with CS dependence structure ($D1$ii), when the longitudinal design is sparse (left) and moderately sparse (right); the functional covariates are observed with high noise variance ($B1$) {and effect $E1$}. Fitted model assumes CS covariance structure. Reported is $RMPE_{trj}$ for observed (white boxplot) and unobserved (gray boxplot) subjects based on 1000 simulations. Reference lines are drawn at RMPE values 1, 2, and 3 for convenience.} 
	\label{trj}
	\noindent\makebox[\textwidth]{	
		\begin{tabular}{cc}
			%\hspace{-.9cm} 
			\includegraphics[angle = 0, height = 7cm, width = 9.2cm]{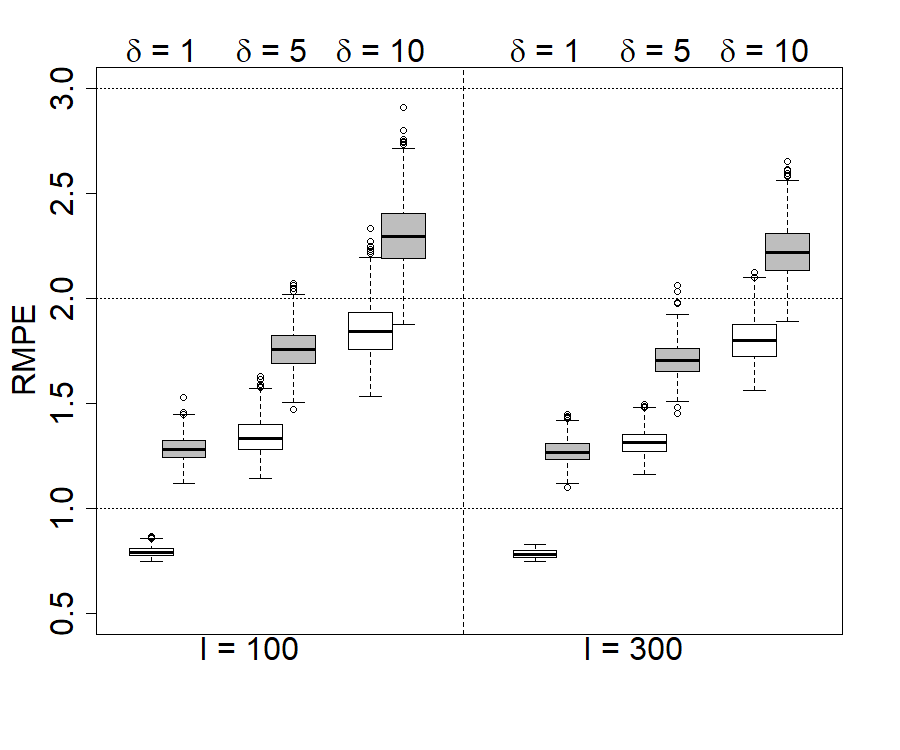}&
			\hspace{-0.75cm}\includegraphics[angle = 0, height = 7cm, width = 9.2cm]{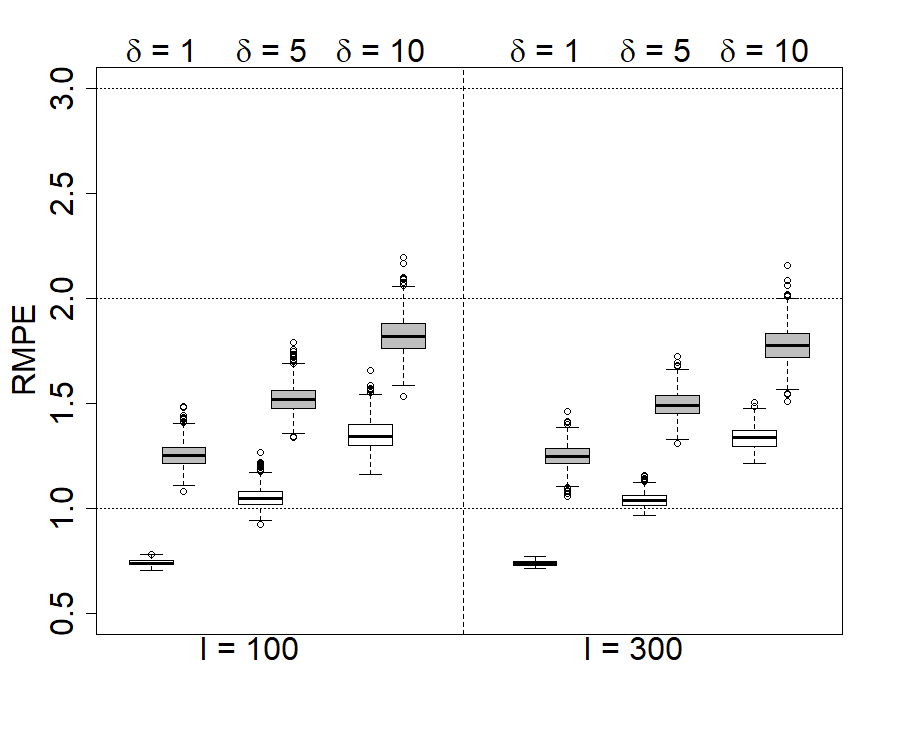}
			%\vspace{-0.5cm}
		\end{tabular}
	}
\end{figure}

\emph{Prediction accuracy of response trajectory}. 
Figure \ref{trj} shows the prediction error ($RMPE_{trj}$) for the entire trajectory in the case of Gaussian responses that are correlated using CS structure ($D2$ii) and are observed in each of the two sampling designs considered, sparse and moderately sparse. As expected, the accuracy improves both as the number of repeated measurements per subject increases and when the sample size increases, the former factor having higher impact. %We also note that the accuracy decreases as $\delta$ departs from zero, essentially as the regression coefficient departs from a time-invariant function. This finding is in agreement with the results shown in Tables  \ref{CS} and\ref{lpeer}, which compare the prediction performance of our approach with the existing alternatives. 
As the magnitude of $\delta$ increases, the difficulty of the problem increases and the prediction accuracy for all cases suffers.

\begin{table}[H]
	\tiny
	\centering
	\caption{Gaussian responses with CS dependence structure ($D1$ii), when the longitudinal design is sparse ($C1$) and moderately sparse (mod sparse, $C2$); the functional covariates are observed with high noise variance ($B1$) {and effect $E1$}. Model is fitted assuming CS type dependence structure. Median prediction errors and IQR in parenthesis are reported for 1000 simulations.}
	\label{CS}
	\noindent\makebox[\textwidth]{ % used to make it centered
		\begin{tabular}{clccccccccclcccc}
			\multicolumn{1}{l}{} &  & \multicolumn{1}{l}{} & \multicolumn{1}{l}{} & \multicolumn{1}{l}{} & \multicolumn{1}{l}{} & \multicolumn{1}{l}{} & \multicolumn{1}{l}{} & \multicolumn{1}{l}{} & \multicolumn{1}{l}{} & \multicolumn{1}{l}{} &  & \multicolumn{1}{l}{} & \multicolumn{1}{l}{} & \multicolumn{1}{l}{} & \multicolumn{1}{l}{} \\
			& \multicolumn{1}{c}{} & \multicolumn{4}{c}{$\delta=0$} & \textbf{} & \multicolumn{4}{c}{$\delta=2$} & \textbf{} & \multicolumn{4}{c}{$\delta=5$} \\ \cline{3-16} 
			{\color[HTML]{000000} \textbf{}} & \multicolumn{1}{c}{{\color[HTML]{6665CD} }} & \multicolumn{2}{c}{{\color[HTML]{000000} $IN_{PE}$}} & \multicolumn{2}{c}{{\color[HTML]{000000} $OUT_{PE}$}} & \textbf{} & \multicolumn{2}{c}{{\color[HTML]{000000} $IN_{PE}$}} & \multicolumn{2}{c}{$OUT_{PE}$} &  & \multicolumn{2}{c}{$IN_{PE}$} & \multicolumn{2}{c}{$OUT_{PE}$} \\ \cline{3-16} 
			& \multicolumn{1}{c}{} & LDFR & LPFR & LDFR & LPFR & \textit{} & LDFR & LPFR & LDFR & LPFR &  & LDFR & LPFR & LDFR & LPFR \\ \cline{3-16} 
			& \multicolumn{1}{c}{} &  &  &  &  &  &  &  &  &  &  & \multicolumn{1}{l}{} & \multicolumn{1}{l}{} & \multicolumn{1}{l}{} & \multicolumn{1}{l}{} \\ \cline{3-16} 
			& sparse & \begin{tabular}[c]{@{}c@{}}0.77\\ (0.04)\end{tabular} & \begin{tabular}[c]{@{}c@{}}0.89\\ (0.04)\end{tabular} & \begin{tabular}[c]{@{}c@{}}0.98\\ (0.11)\end{tabular} & \begin{tabular}[c]{@{}c@{}}1.00\\ (0.05)\end{tabular} &  & \begin{tabular}[c]{@{}c@{}}0.87\\ (0.05)\end{tabular} & \begin{tabular}[c]{@{}c@{}}1.37\\ (0.08)\end{tabular} & \begin{tabular}[c]{@{}c@{}}1.11\\ (0.11)\end{tabular} & \begin{tabular}[c]{@{}c@{}}1.53\\ (0.09)\end{tabular} &  & \begin{tabular}[c]{@{}c@{}}1.30\\ (0.13)\end{tabular} & \begin{tabular}[c]{@{}c@{}}3.32\\ (0.30)\end{tabular} & \begin{tabular}[c]{@{}c@{}}1.73\\ (0.25)\end{tabular} & \begin{tabular}[c]{@{}c@{}}3.54\\ (0.32)\end{tabular} \\ \cline{3-16} 
			{I = 100} & mod sparse & \begin{tabular}[c]{@{}c@{}}0.74\\ (0.03)\end{tabular} & \begin{tabular}[c]{@{}c@{}}0.91\\ (0.03)\end{tabular} & \begin{tabular}[c]{@{}c@{}}0.81\\ (0.05)\end{tabular} & \begin{tabular}[c]{@{}c@{}}0.96\\ (0.04)\end{tabular} &  & \begin{tabular}[c]{@{}c@{}}0.79\\ (0.03)\end{tabular} & \begin{tabular}[c]{@{}c@{}}1.40\\ (0.06)\end{tabular} & \begin{tabular}[c]{@{}c@{}}0.86\\ (0.06)\end{tabular} & \begin{tabular}[c]{@{}c@{}}1.48\\ (0.09)\end{tabular} &  & \begin{tabular}[c]{@{}c@{}}1.02\\ (0.08)\end{tabular} & \begin{tabular}[c]{@{}c@{}}3.36\\ (0.26)\end{tabular} & \begin{tabular}[c]{@{}c@{}}1.15\\ (0.12)\end{tabular} & \begin{tabular}[c]{@{}c@{}}3.49\\ (0.32)\end{tabular} \\ \cline{3-16} 
			\textbf{} & \multicolumn{1}{c}{\textbf{}} &  &  &  &  &  &  &  &  &  &  &  &  &  &  \\ \cline{3-16} 
			& sparse & \begin{tabular}[c]{@{}c@{}}0.76\\ (0.03)\end{tabular} & \begin{tabular}[c]{@{}c@{}}0.88\\ (0.03)\end{tabular} & \begin{tabular}[c]{@{}c@{}}0.94\\ (0.09)\end{tabular} & \begin{tabular}[c]{@{}c@{}}1.00\\ (0.04)\end{tabular} &  & \begin{tabular}[c]{@{}c@{}}0.85\\ (0.03)\end{tabular} & \begin{tabular}[c]{@{}c@{}}1.37\\ (0.06)\end{tabular} & \begin{tabular}[c]{@{}c@{}}1.05\\ (0.08)\end{tabular} & \begin{tabular}[c]{@{}c@{}}1.53\\ (0.07)\end{tabular} &  & \begin{tabular}[c]{@{}c@{}}1.26\\ (0.09)\end{tabular} & \begin{tabular}[c]{@{}c@{}}3.34\\ (0.22)\end{tabular} & \begin{tabular}[c]{@{}c@{}}1.60\\ (0.17)\end{tabular} & \begin{tabular}[c]{@{}c@{}}3.54\\ (0.23)\end{tabular} \\ \cline{3-16} 
			{I = 200} & mod sparse & \begin{tabular}[c]{@{}c@{}}0.73\\ (0.02)\end{tabular} & \begin{tabular}[c]{@{}c@{}}0.90\\ (0.02)\end{tabular} & \begin{tabular}[c]{@{}c@{}}0.79\\ (0.04)\end{tabular} & \begin{tabular}[c]{@{}c@{}}0.95\\ (0.03)\end{tabular} &  & \begin{tabular}[c]{@{}c@{}}0.78\\ (0.02)\end{tabular} & \begin{tabular}[c]{@{}c@{}}1.41\\ (0.05)\end{tabular} & \begin{tabular}[c]{@{}c@{}}0.84\\ (0.04)\end{tabular} & \begin{tabular}[c]{@{}c@{}}1.48\\ (0.07)\end{tabular} &  & \begin{tabular}[c]{@{}c@{}}0.99\\ (0.05)\end{tabular} & \begin{tabular}[c]{@{}c@{}}3.38\\ (0.18)\end{tabular} & \begin{tabular}[c]{@{}c@{}}1.09\\ (0.08)\end{tabular} & \begin{tabular}[c]{@{}c@{}}3.50\\ (0.25)\end{tabular} \\ \cline{3-16} 
			\multicolumn{1}{l}{} &  & \multicolumn{1}{l}{} & \multicolumn{1}{l}{} & \multicolumn{1}{l}{} & \multicolumn{1}{l}{} & \multicolumn{1}{l}{} & \multicolumn{1}{l}{} & \multicolumn{1}{l}{} & \multicolumn{1}{l}{} & \multicolumn{1}{l}{} &  & \multicolumn{1}{l}{} & \multicolumn{1}{l}{} & \multicolumn{1}{l}{} & \multicolumn{1}{l}{} \\ \cline{3-16} 
			\multicolumn{1}{l}{} & sparse & \begin{tabular}[c]{@{}c@{}}0.76\\ (0.03)\end{tabular} & \begin{tabular}[c]{@{}c@{}}0.88\\ (0.02)\end{tabular} & \begin{tabular}[c]{@{}c@{}}0.91\\ (0.08)\end{tabular} & \begin{tabular}[c]{@{}c@{}}0.99\\ (0.03)\end{tabular} &  & \begin{tabular}[c]{@{}c@{}}0.84\\ (0.03)\end{tabular} & \begin{tabular}[c]{@{}c@{}}1.37\\ (0.05)\end{tabular} & \begin{tabular}[c]{@{}c@{}}1.02\\ (0.06)\end{tabular} & \begin{tabular}[c]{@{}c@{}}1.53\\ (0.05)\end{tabular} &  & \begin{tabular}[c]{@{}c@{}}1.24\\ (0.07)\end{tabular} & \begin{tabular}[c]{@{}c@{}}3.35\\ (0.18)\end{tabular} & \begin{tabular}[c]{@{}c@{}}1.54\\ (0.15)\end{tabular} & \begin{tabular}[c]{@{}c@{}}3.54\\ (0.19)\end{tabular} \\ \cline{3-16} 
			\multicolumn{1}{l}{I = 300} & mod sparse & \begin{tabular}[c]{@{}c@{}}0.73\\ (0.02)\end{tabular} & \begin{tabular}[c]{@{}c@{}}0.90\\ (0.01)\end{tabular} & \begin{tabular}[c]{@{}c@{}}0.79\\ (0.04)\end{tabular} & \begin{tabular}[c]{@{}c@{}}0.95\\ (0.02)\end{tabular} &  & \begin{tabular}[c]{@{}c@{}}0.77\\ (0.02)\end{tabular} & \begin{tabular}[c]{@{}c@{}}1.41\\ (0.04)\end{tabular} & \begin{tabular}[c]{@{}c@{}}0.83\\ (0.03)\end{tabular} & \begin{tabular}[c]{@{}c@{}}1.48\\ (0.05)\end{tabular} &  & \begin{tabular}[c]{@{}c@{}}0.98\\ (0.04)\end{tabular} & \begin{tabular}[c]{@{}c@{}}3.38\\ (0.15)\end{tabular} & \begin{tabular}[c]{@{}c@{}}1.07\\ (0.06)\end{tabular} & \begin{tabular}[c]{@{}c@{}}3.49\\ (0.20)\end{tabular} \\ \cline{3-16} 
			\multicolumn{1}{l}{} &  & \multicolumn{1}{l}{} & \multicolumn{1}{l}{} & \multicolumn{1}{l}{} & \multicolumn{1}{l}{} & \multicolumn{1}{l}{} & \multicolumn{1}{l}{} & \multicolumn{1}{l}{} & \multicolumn{1}{l}{} & \multicolumn{1}{l}{} &  & \multicolumn{1}{l}{} & \multicolumn{1}{l}{} & \multicolumn{1}{l}{} & \multicolumn{1}{l}{}
		\end{tabular}
	}
\end{table}

\begin{table}[H]
	\tiny
	\centering
	\caption{Gaussian responses with CS dependence structure ($D1$ii), when the longitudinal design is sparse ($C1$) for $B1$ and $B2$ {with effect $E1$}. Model is fitted assuming CS type dependence structure. Median prediction errors and IQR in parenthesis are reported.}
	\label{lpeer}
	\noindent\makebox[\textwidth]{ % used to make it centered
	\begin{tabular}{cccccccclccclcc}
		\multicolumn{1}{l}{} & \multicolumn{1}{l}{} & \multicolumn{1}{l}{} & \multicolumn{1}{l}{} & \multicolumn{1}{l}{} & \multicolumn{1}{l}{} & \multicolumn{1}{l}{} & \multicolumn{1}{l}{} &  & \multicolumn{1}{l}{} & \multicolumn{1}{l}{} & \multicolumn{1}{l}{} &  & \multicolumn{1}{l}{} & \multicolumn{1}{l}{} \\
		&  & \multicolumn{6}{c}{$\delta=1$} &  & \multicolumn{6}{c}{$\delta=5$} \\ \cline{3-15} 
		{\color[HTML]{000000} \textbf{}} & {\color[HTML]{6665CD} } & \multicolumn{2}{c}{{\color[HTML]{000000} $IN_{PE}$}} & \multicolumn{2}{c}{{\color[HTML]{000000} $OUT_{PE}$}} & \multicolumn{2}{c}{Run time ('sec)} &  & \multicolumn{2}{c}{{\color[HTML]{000000} $IN_{PE}$}} & \multicolumn{2}{c}{$OUT_{PE}$} & \multicolumn{2}{c}{Run time ('sec)} \\ \cline{3-15} 
		& \textit{} & LDFR & LPEER & LDFR & LPEER & LDFR & LPEER & \textit{} & LDFR & LPEER & LDFR & LPEER & LDFR & LPEER \\ \cline{3-15} 
		&  &  &  &  &  &  &  &  &  &  &  &  &  &  \\ \cline{3-15} 
		& SNR 0.5 & \begin{tabular}[c]{@{}c@{}}0.74\\ (0.03)\end{tabular} & \begin{tabular}[c]{@{}c@{}}0.82\\ (0.03)\end{tabular} & \begin{tabular}[c]{@{}c@{}}0.91\\ (0.07)\end{tabular} & \begin{tabular}[c]{@{}c@{}}0.94\\ (0.05)\end{tabular} & 14.74 & 380.75 & \multicolumn{1}{c}{} & \begin{tabular}[c]{@{}c@{}}1.31\\ (0.13)\end{tabular} & \begin{tabular}[c]{@{}c@{}}1.78\\ (0.15)\end{tabular} & \begin{tabular}[c]{@{}c@{}}1.72\\ (0.24)\end{tabular} & \multicolumn{1}{c}{\begin{tabular}[c]{@{}c@{}}2.05\\ (0.15)\end{tabular}} & 10.60 & 703.65 \\ \cline{3-15} 
		{I = 100} & SNR 2.5 & \begin{tabular}[c]{@{}c@{}}0.73\\ (0.03)\end{tabular} & \begin{tabular}[c]{@{}c@{}}0.79\\ (0.03)\end{tabular} & \begin{tabular}[c]{@{}c@{}}0.90\\ (0.07)\end{tabular} & \begin{tabular}[c]{@{}c@{}}0.91\\ (0.03)\end{tabular} & 12.92 & 210.95 & \multicolumn{1}{c}{} & \begin{tabular}[c]{@{}c@{}}1.30\\ (0.14)\end{tabular} & \begin{tabular}[c]{@{}c@{}}1.73\\ (0.12)\end{tabular} & \begin{tabular}[c]{@{}c@{}}1.75\\ (0.28)\end{tabular} & \multicolumn{1}{c}{\begin{tabular}[c]{@{}c@{}}2.00\\ (0.14)\end{tabular}} & 11.28 & 805.15 \\ \cline{3-15} 
		\textbf{} & \textbf{} &  &  &  &  &  &  &  &  &  &  &  &  &  \\ \cline{3-15} 
		& SNR 0.5 & \begin{tabular}[c]{@{}c@{}}0.73\\ (0.02)\end{tabular} & \begin{tabular}[c]{@{}c@{}}0.81\\ (0.02)\end{tabular} & \begin{tabular}[c]{@{}c@{}}0.86\\ (0.05)\end{tabular} & \begin{tabular}[c]{@{}c@{}}0.92\\ (0.03)\end{tabular} & 97.00 & 1270.51 & \multicolumn{1}{c}{} & \begin{tabular}[c]{@{}c@{}}1.25\\ (0.06)\end{tabular} & \begin{tabular}[c]{@{}c@{}}1.78\\ (0.07)\end{tabular} & \begin{tabular}[c]{@{}c@{}}1.56\\ (0.13)\end{tabular} & \multicolumn{1}{c}{\begin{tabular}[c]{@{}c@{}}2.02\\ (0.07)\end{tabular}} & 63.75 & 4542.27 \\ \cline{3-15} 
		{I = 300} & SNR 2.5 & \begin{tabular}[c]{@{}c@{}}0.72\\ (0.02)\end{tabular} & \begin{tabular}[c]{@{}c@{}}0.80\\ (0.02)\end{tabular} & \begin{tabular}[c]{@{}c@{}}0.85\\ (0.04)\end{tabular} & \begin{tabular}[c]{@{}c@{}}0.90\\ (0.03)\end{tabular} & 90.30 & 1230.92 & \multicolumn{1}{c}{} & \begin{tabular}[c]{@{}c@{}}1.20\\ (0.09)\end{tabular} & \begin{tabular}[c]{@{}c@{}}1.71\\ (0.06)\end{tabular} & \begin{tabular}[c]{@{}c@{}}1.51\\ (0.15)\end{tabular} & \multicolumn{1}{c}{\begin{tabular}[c]{@{}c@{}}1.96\\ (0.16)\end{tabular}} & 71.55 & 4367.22 \\ \cline{3-15} 
		\multicolumn{1}{l}{} & \multicolumn{1}{l}{} & \multicolumn{1}{l}{} & \multicolumn{1}{l}{} & \multicolumn{1}{l}{} & \multicolumn{1}{l}{} & \multicolumn{1}{l}{} & \multicolumn{1}{l}{} &  & \multicolumn{1}{l}{} & \multicolumn{1}{l}{} & \multicolumn{1}{l}{} &  & \multicolumn{1}{l}{} & \multicolumn{1}{l}{}
	\end{tabular}
}
\end{table}

\begin{table}[H]
	\tiny
	\centering
	\caption{Binary responses ($D2$) when the longitudinal design is sparse/moderately sparse (mod sparse); the functional covariates are observed with high noise variance ($B1$) {and effect $E1$}. Fitted model assumes subject-specific random intercept. We report the median of prediction errors of the linear predictor trajectories and, in parenthesis, the median of true-positive-rates.}	
	\label{binarySNR5_larger_noise}
	\noindent\makebox[\textwidth]{ % used to make it centered
		\begin{tabular}{llccccccccc}
			&  & \multicolumn{1}{l}{} & \multicolumn{1}{l}{} & \multicolumn{1}{l}{} & \multicolumn{1}{l}{} & \multicolumn{1}{l}{} & \multicolumn{1}{l}{} & \multicolumn{1}{l}{} & \multicolumn{1}{l}{} & \multicolumn{1}{l}{} \\
			&  & \multicolumn{4}{c}{$\delta = 0$} &  & \multicolumn{4}{c}{$\delta = 5$} \\ \cline{3-11} 
			&  & \multicolumn{2}{c}{$IN_{PE}$} & \multicolumn{2}{c}{$OUT_{PE}$} &  & \multicolumn{2}{c}{$IN_{PE}$} & \multicolumn{2}{c}{$OUT_{PE}$} \\ \cline{3-11} 
			&  & LDFR & LPFR & LDFR & LPFR &  & LDFR & LPFR & LDFR & LPFR \\ \cline{3-11} 
			&  & \multicolumn{1}{l}{} & \multicolumn{1}{l}{} & \multicolumn{1}{l}{} & \multicolumn{1}{l}{} & \multicolumn{1}{l}{} & \multicolumn{1}{l}{} & \multicolumn{1}{l}{} & \multicolumn{1}{l}{} & \multicolumn{1}{l}{} \\ \cline{3-11} 
			I = 100 & sparse & \begin{tabular}[c]{@{}c@{}}1.27\\ (0.96)\end{tabular} & \begin{tabular}[c]{@{}c@{}}1.30\\ (0.96)\end{tabular} & \begin{tabular}[c]{@{}c@{}}1.33\\ (0.96)\end{tabular} & \begin{tabular}[c]{@{}c@{}}1.30\\ (0.96)\end{tabular} &  & \begin{tabular}[c]{@{}c@{}}1.89\\ (0.90)\end{tabular} & \begin{tabular}[c]{@{}c@{}}2.88\\ (0.87)\end{tabular} & \begin{tabular}[c]{@{}c@{}}2.07\\ (0.90)\end{tabular} & \begin{tabular}[c]{@{}c@{}}2.90\\ (0.86)\end{tabular} \\ \cline{3-11} 
			& mod sparse & \begin{tabular}[c]{@{}c@{}}1.19\\ (0.96)\end{tabular} & \begin{tabular}[c]{@{}c@{}}1.29\\ (0.96)\end{tabular} & \begin{tabular}[c]{@{}c@{}}1.20\\ (0.96)\end{tabular} & \begin{tabular}[c]{@{}c@{}}1.29\\ (0.96)\end{tabular} &  & \begin{tabular}[c]{@{}c@{}}1.55\\ (0.91)\end{tabular} & \begin{tabular}[c]{@{}c@{}}2.87\\ (0.86)\end{tabular} & \begin{tabular}[c]{@{}c@{}}1.59\\ (0.91)\end{tabular} & \begin{tabular}[c]{@{}c@{}}2.87\\ (0.86)\end{tabular} \\ \cline{3-11} 
			&  & \multicolumn{1}{l}{} & \multicolumn{1}{l}{} & \multicolumn{1}{l}{} & \multicolumn{1}{l}{} & \multicolumn{1}{l}{} & \multicolumn{1}{l}{} & \multicolumn{1}{l}{} & \multicolumn{1}{l}{} & \multicolumn{1}{l}{} \\ \cline{3-11} 
			I = 200 & sparse & \begin{tabular}[c]{@{}c@{}}1.23\\ (0.96)\end{tabular} & \begin{tabular}[c]{@{}c@{}}1.29\\ (0.96)\end{tabular} & \begin{tabular}[c]{@{}c@{}}1.28\\ (0.96)\end{tabular} & \begin{tabular}[c]{@{}c@{}}1.29\\ (0.96)\end{tabular} &  & \begin{tabular}[c]{@{}c@{}}1.81\\ (0.90)\end{tabular} & \begin{tabular}[c]{@{}c@{}}2.89\\ (0.86)\end{tabular} & \begin{tabular}[c]{@{}c@{}}1.95\\ (0.90)\end{tabular} & \begin{tabular}[c]{@{}c@{}}2.90\\ (0.86)\end{tabular} \\ \cline{3-11} 
			& mod sparse & \begin{tabular}[c]{@{}c@{}}1.17\\ (0.96)\end{tabular} & \begin{tabular}[c]{@{}c@{}}1.28\\ (0.96)\end{tabular} & \begin{tabular}[c]{@{}c@{}}1.18\\ (0.96)\end{tabular} & \begin{tabular}[c]{@{}c@{}}1.29\\ (0.96)\end{tabular} &  & \begin{tabular}[c]{@{}c@{}}1.46\\ (0.92)\end{tabular} & \begin{tabular}[c]{@{}c@{}}2.87\\ (0.86)\end{tabular} & \begin{tabular}[c]{@{}c@{}}1.49\\ (0.91)\end{tabular} & \begin{tabular}[c]{@{}c@{}}2.88\\ (0.86)\end{tabular} \\ \cline{3-11} 
			&  & \multicolumn{1}{l}{} & \multicolumn{1}{l}{} & \multicolumn{1}{l}{} & \multicolumn{1}{l}{} & \multicolumn{1}{l}{} & \multicolumn{1}{l}{} & \multicolumn{1}{l}{} & \multicolumn{1}{l}{} & \multicolumn{1}{l}{} \\ \cline{3-11} 
			I = 300 & sparse & \begin{tabular}[c]{@{}c@{}}1.21\\ (0.96)\end{tabular} & \begin{tabular}[c]{@{}c@{}}1.28\\ (0.96)\end{tabular} & \begin{tabular}[c]{@{}c@{}}1.25\\ (0.96)\end{tabular} & \begin{tabular}[c]{@{}c@{}}1.29\\ (0.96)\end{tabular} &  & \begin{tabular}[c]{@{}c@{}}1.77\\ (0.91)\end{tabular} & \begin{tabular}[c]{@{}c@{}}2.89\\ (0.87)\end{tabular} & \begin{tabular}[c]{@{}c@{}}1.88\\ (0.91)\end{tabular} & \begin{tabular}[c]{@{}c@{}}2.89\\ (0.86)\end{tabular} \\ \cline{3-11} 
			& mod sparse & \begin{tabular}[c]{@{}c@{}}1.17\\ (0.96)\end{tabular} & \begin{tabular}[c]{@{}c@{}}1.28\\ (0.96)\end{tabular} & \begin{tabular}[c]{@{}c@{}}1.17\\ (0.96)\end{tabular} & \begin{tabular}[c]{@{}c@{}}1.28\\ (0.96)\end{tabular} &  & \begin{tabular}[c]{@{}c@{}}1.43\\ (0.92)\end{tabular} & \begin{tabular}[c]{@{}c@{}}2.88\\ (0.86)\end{tabular} & \begin{tabular}[c]{@{}c@{}}1.45\\ (0.92)\end{tabular} & \begin{tabular}[c]{@{}c@{}}2.88\\ (0.86)\end{tabular} \\ \cline{3-11} 
			&  & \multicolumn{1}{l}{} & \multicolumn{1}{l}{} & \multicolumn{1}{l}{} & \multicolumn{1}{l}{} & \multicolumn{1}{l}{} & \multicolumn{1}{l}{} & \multicolumn{1}{l}{} & \multicolumn{1}{l}{} & \multicolumn{1}{l}{}
		\end{tabular}
	}
\end{table}

%
%\begin{figure}[H] 
%	\caption{Gaussian responses with CS dependence structure ($D1$ii), when the longitudinal design is sparse (top) and moderately sparse (bottom); the functional covariates are observed with high noise variance ($B1$) {and effect $E1$}. Fitted model assumes CS covariance structure. Reported is $RMPE_{trj}$ based on 1000 simulations.}
%	\label{trj}
%	\noindent\makebox[\textwidth]{
%	\includegraphics[width=0.80\textwidth]{lowSNR.png}
%}
%\end{figure} 
%

\emph{Accuracy of the prediction bands}. We examine the performance of the prediction bands in terms of actual coverage and expected length for two nominal levels $90\%$ and $95\%$. Table \ref{Gaussian_coverage} shows the results for few choices of $\delta$. In general the average coverage stays around the nominal levels for both the observed and unobserved subjects across different settings, and irrespective of the complexity of the signal as defined by $\delta$; see the results for $\delta=1$ and $\delta=10$. As expected, the width of the prediction bands is larger for the new subjects compared to the existing one and it increases with the complexity of the regression coefficient (large $|\delta|$). For all the settings, the results improve for larger sample size and increased signal to noise ratio (SNR).

\begin{table}[H]
	\tiny
	\centering
	\caption{Gaussian responses with CS dependence structure ($D1$ii), when the longitudinal design is moderately sparse; the functional covariates are observed with high ($B1$) and low ($B2$) noise variance with effect $E1$. Fitted model assumes CS covariance structure. Reported are the average coverage probabilities of 95\% and 90\% pointwise prediction bands, standard errors (in parenthesis), average length (in square bracket), for the observed ($Y_{i}$) and unobserved subjects ($Y_{i^{*}}$). Results are based on 1000 MC simulations.}
	\label{Gaussian_coverage}
	\noindent\makebox[\textwidth]{
	\begin{tabular}{lllccccccc}
		&  & \multicolumn{1}{c}{} & \multicolumn{3}{c}{(1 - $\alpha$) = 0.95} &  & \multicolumn{3}{c}{(1 - $\alpha$) = 0.90} \\ \cline{4-6} \cline{8-10} 
		&  & \multicolumn{1}{c}{} & $\delta = 1$ & $\delta = 5$ & $\delta = 10$ &  & $\delta = 1$ & $\delta = 5$ & $\delta = 10$ \\ \cline{4-6} \cline{8-10} 
		{SNR = 0.5} & {I = 100} & $Y_{i}$ & 0.96 (0.01) {[}2.99{]} & 0.95 (0.01) {[}4.18{]} & 0.95 (0.01) {[}5.39{]} &  & 0.91 (0.01) {[}2.51{]} & 0.92 (0.01) {[}3.51{]} & 0.91 (0.01) {[}4.53{]} \\ \cline{4-6} \cline{8-10} 
		&  & $Y_{i^{*}}$ & 0.95 (0.01) {[}4.91{]} & 0.94 (0.01) {[}5.85{]} & 0.94 (0.01) {[}6.96{]} &  & 0.90 (0.01) {[}4.12{]} & 0.90 (0.01) {[}4.91{]} & 0.90 (0.01) {[}5.84{]} \\ \cline{4-6} \cline{8-10} 
		&  & \multicolumn{1}{c}{} &  &  &  &  &  &  &  \\ \cline{4-6} \cline{8-10} 
		& {I = 300} & $Y_{i}$ & 0.95 (0.01) {[}2.97{]} & 0.95 (0.01) {[}4.12{]} & 0.95 (0.01) {[}5.30{]} &  & 0.91 (0.01) {[}2.49{]} & 0.91 (0.01) {[}3.46{]} & 0.91 (0.01) {[}4.44{]} \\ \cline{4-6} \cline{8-10} 
		&  & $Y_{i^{*}}$ & 0.95 (0.01) {[}4.89{]} & 0.95 (0.01) {[}5.77{]} & 0.94 (0.01) {[}6.86{]} &  & 0.90 (0.01) {[}4.11{]} & 0.90 (0.01) {[}4.84{]} & 0.90 (0.01) {[}5.76{]} \\ \cline{4-6} \cline{8-10} 
		&  &  &  &  &  &  &  &  &  \\ \cline{4-6} \cline{8-10} 
		{SNR = 2.5} & {I = 100} & $Y_{i}$ & 0.96 (0.01) {[}2.97{]} & 0.95 (0.01) {[}4.07{]} & 0.95 (0.01) {[}5.20{]} &  & 0.91 (0.01) {[}2.49{]} & 0.92 (0.01) {[}3.41{]} & 0.91 (0.01) {[}4.36{]} \\ \cline{4-6} \cline{8-10} 
		&  & $Y_{i^{*}}$ & 0.95 (0.01) {[}4.91{]} & 0.94 (0.01) {[}5.75{]} & 0.94 (0.01) {[}6.80{]} &  & 0.90 (0.01) {[}4.12{]} & 0.90 (0.01) {[}4.83{]} & 0.90 (0.01) {[}5.71{]} \\ \cline{4-6} \cline{8-10} 
		&  &  &  &  &  &  &  &  &  \\ \cline{4-6} \cline{8-10} 
		& {I = 300} & $Y_{i}$ & 0.96 (0.01) {[}2.95{]} & 0.95 (0.01) {[}4.01{]} & 0.95 (0.01) {[}5.08{]} &  & 0.91 (0.01) {[}2.47{]} & 0.91 (0.01) {[}3.37{]} & 0.91 (0.01) {[}4.27{]} \\ \cline{4-6} \cline{8-10} 
		&  & $Y_{i^{*}}$ & 0.95 (0.01) {[}4.89{]} & 0.95 (0.01) {[}5.67{]} & 0.94 (0.01) {[}6.67{]} &  & 0.90 (0.01) {[}4.10{]} & 0.90 (0.01) {[}4.77{]} & 0.90 (0.01) {[}6.00{]} \\ \cline{4-6} \cline{8-10} 
	\end{tabular}
}
\end{table}

\section{Data application} \label{sec:data_application}

Our motivating application is a lactating sow study where the primary objective is to investigate the effect of thermal environment (i.e. temperature ($T$)) on the feed-intake of the lactating sows. This study is very important for several reasons: (1) ambient temperatures above the evaporative critical temperature decreases the amount of food-intake which, as a result, deteriorates the reproductive performance and hinders the growth rate of piglets of lactating sows (\cite{black1993lactation}. (2) Also, poor feed-intake of the lactating sows 
leads to increased body weight loss during lactation and reduced milk yield, and is further associated with compromised weight gain of their litter (\cite{johnston1999effect}, \cite{renaudeau2001effects}, \cite{renaudeau2001effects}). (3) Thirdly, heat-stress results in a reduction of farrowing rate (the percentage of sows that become pregnant and farrow a litter of piglets) and total number of pigs born in sows (\cite{bloemhof2013effect}) which in turn has a negative effect on the total production of pork meat per year. (4) Fourth, pigs from sows raised in an unfavorable thermal environment will be fatter than the ones reared in favorable cooler environments and this fact makes pork meat fattier (\cite{BaumgardL2015}). (5) Fifth, because of heat stress associated with hot climactic thermal environment, the swine industry in US incurs a total estimated loss worth of $\$300$ million per year on average (\cite{st2003economic}). Therefore, insight into how the feeding behavior changes over time due to the prolonged exposure to a hot environment will assist in proposing more economical and efficient feeding strategies for lactating sows.

The experimental study was carried during July to October in 2013 in a 2,600-sow commercial research unit in Oklahoma (\cite{rosero2016essential}) and involved 480 PIC Camborough sows. The sows were kept in the farrowing facility where they gave birth to piglets. Depending on the number of previous pregnancies (parity levels), sows were classified into younger (parity equal to zero or one) or older (parity equal to two or higher). The sows are brought to the farrowing crates when they are approximately five days before they are due to give birth. They arrive in groups; the study involves 21 groups of about 21-23 sows.
Sows are observed during their 20-21 day lactation period and their respective food-consumption is monitored. Each sow is provided food individually with a computerized feeding system (Howema, Big Dutchman, Germany). The amount of food-offered $(FO)$ was recorded at 2.00PM on each day and feed-refusal $(FR)$ was measured the following day prior to any subsequent food addition; feed-intake $(FI)$ was calculated as $FI = FO- FR$ in kg. Minute-by-minute information about the ambient air temperature (${^\circ}C$) and humidity ($\%$) of the farrowing facility were recorded by data loggers (LogTag, MicroDAQ Ltd., Contoocook, NH). The experimenters removed information of five sows due to unreliable measurements and thus we had available information for 475 sows. The facility ambient temperature was controlled by a ventilation system; the barns have cool cells that pull fresh air through wet corrugated material to provide further cooling of air. There are some missing observations for temperature profiles due to machine failure which qualifies the pattern of missingness as missing completely at random. %\textcolor{red}{WE MAY DELETE THIS Preliminary investigation confirms that the ambient temperature (dry-bulb temperature) and humidity are negatively correlated (see also \cite{lawrence2005relationship}).} Thus, we focus solely on the ambient temperature in this paper; 
Our objective is to study the effect of temperature on the feed intake of sows.

Let $i$ index the sows, $j$ index the repeated instances for the same sow, and $t_{ij}$ to denote the lactation day of the $i$th sow, corresponding to the $j$th instance at which the sow is observed; for many sows we have $t_{ij}=j$ for $j=1, \ldots 21$, but this is not always the case. The ``time" is defined as the 24 hours time window from 2:00PM to 1:59PM. Furthermore let $g_i$ index the group of the $i$th sow, $g_i=1, \ldots, 21$; typically, the sows within the same group give birth closer to one another. Denote by $nTemp_{ij}(\cdot) = nTemp_{i}(\cdot, t_{ij})$ the daily {temperature} profile observed at the $t_{ij}$ lactation day of the $i$th sow; the measurements typically include noise, hence the prefix ``n". Later we use notation $Temp_{ij}(\cdot)$ for the true {temperature} profile corresponding to $nTemp_{ij}(\cdot)$.  The right panel of Figure \ref{Sow14473} shows the daily {temperature} corresponding to the first 21st days for a random sow. Let $FI_{ij}$ be the $FI$ of the $i$th sow at its $j$th repeated occasion, lactation day $t_{ij}$. 

We assume that the relationship between the feed intake and the {temperature} is described by the LDFR model:
\begin{equation}
\label{SOW}
FI_{ij} =   \beta_{p_{i}}(t_{ij}) + \int_{\mathcal{S}} Temp_{ij}(s) \gamma(s,t_{ij}) ds  + b_{g_i} + b_{0i(g_i)} + b_{1i(g_i)}t_{ij} + \varepsilon_{ij},
\end{equation} 
where $\beta_{p_{i}}(\cdot)$ is the mean feed intake for group $p_i$ where $p_i=0$ (young sows) and $p_i=1$ (old sows) and $\gamma(s, \cdot)$ quantifies the time-varying effect of the {temperature} on $FI$; the integral reflects the aggregated effect during the course of the 24 hours. The term $b_{g_i} + b_{0i(g_i)} + b_{1i(g_i)}t_{ij} $ models the dependence of the responses within the same sow as well as the dependence of the responses of the sows who are in the same group. The random terms $ b_{g_i}$ is a group-specific effect and $b_{0i(g_i)}$ and $b_{1i(g_i)}$ are sow within group-specific intercept and slope.  We assume that $b_{g_i}\sim N (0, \sigma_{g}^{2})$, $b_{0i(g_i)} \sim N (0, \sigma_0^{2})$, and $b_{1i(g_i)} \sim N (0, \sigma_1^{2})$ are all mutually independent. Finally, it is assumed that the measurement errors $\varepsilon_{ij}$ are independent and distributed as $N (0, \sigma_{e}^{2})$. Model (\ref{SOW}) does not account for the previous day feed intake, which may be viewed as an important predictor for the current day feed intake. Such approach is discussed later this section.

The steps for fitting the model (\ref{SOW}) are similar to the ones described in Section \ref{sec:simulation}. One important specific is that the fast bivariate spline smoothing (\cite{xiao2013fast}) uses cubic splines with $35$ knots in $s$-direction and $19$ knots in $t$-direction and second order difference penalty is used to control the smoothness of the fit and the tuning parameters are estimated by REML. Also the parameter functions $\beta_{k}(\cdot)$ are modeled using 15 truncated linear splines with knots placed uniformly in the time domain; the smoothing parameters are selected using REML.

\subsection{Fit assessment}
     
We first assess the prediction performance of the proposed method and compare it with the available competitors, LPFR and LPEER. In this regard we split the data into a train set, which is used to build the model and a test set on which the prediction performance is evaluated; {we replicate the test-train split for 25 times}. We consider two ways of forming the test data: (a) randomly select $350$ sows of the total of $475$ sows and include only the measurements corresponding to their last $10$ lactation days; and (b) take all the $475$ sows and include only the measurements corresponding to about $20\%$ of their lactation days that are selected at random. The remaining data form the training set. Approach (a) involves data on fewer sows than approach (b). At the same time, the approach (a) assesses the performance of prediction at ``future" lactation days, while the approach (b) evaluates the prediction performance at random lactation days within the $1$ to $21$ days at which the sows are observed. 

For completeness we describe the implementation of the competitive approaches. LPFR assumes that temperature has a constant effect across the lactation days of a sow and models this time-invariant effect using $30$ truncated linear splines basis functions and the tuning parameters are estimated by REML. The covariance structure is specified as in model (\ref{SOW}) and the model is fitted using {\tt lpfr()} function available in {\tt refund} package (\cite{refund}). For LPEER, we consider polynomial functions of time of degree $d=0,1,\ldots, 4$, $\gamma(s,t) = \gamma_1(s) + t \gamma_1(s) + \cdots + t^d \gamma_d(s)$ and select the optimal $d$ by AIC as described earlier; the model is fitted using {\tt lpeer} in the same package. The covariance structure is specified using a subject-specific random intercept and random group effect; it is not clear how to modify the existing code to accommodate a subject-specific slope effect.  

Table \ref{comparison_sow_temperature} reports the results. The findings show that LDFR and LPEER perform relatively similar in terms of in-sample and out-of-sample prediction in the two situations considered. Nevertheless our methodology yields to computations that are an order of magnitude faster and a better fitting criteria as assessed by AIC. In contrast, the prediction results with LPFR are inferior, and they indicate a lack of appropriateness of a time-invariant effect model for our lactating sow application, based on the observations from our simulation investigation.

\begin{table}[H]
	\tiny
	\centering
	\caption{Median prediction accuracy, computing time (in seconds), and marginal AIC for the proposed LDFR and the competitive approaches LPFR and LPEER for the sows application; {corresponding IQRs are reported in parenthesis}.}
   \label{comparison_sow_temperature}
\noindent\makebox[\textwidth]{
	\begin{tabular}{ccccccccccccccccc}
		\cline{3-5} \cline{7-9} \cline{11-13} \cline{15-17}
		&  & \multicolumn{3}{c}{IN$_{PE}$} &  & \multicolumn{3}{c}{OUT$_{PE}$} &  & \multicolumn{3}{c}{AIC} &  & \multicolumn{3}{c}{Computing time} \\ \cline{3-5} \cline{7-9} \cline{11-13} \cline{15-17} 
		&  & LDFR & LPEER & LPFR &  & LDFR & LPEER & LPFR &  & LDFR & LPEER & LPFR &  & LDFR & LPEER & LPFR \\ \cline{3-5} \cline{7-9} \cline{11-13} \cline{15-17} 
		(a) &  & \begin{tabular}[c]{@{}c@{}}1.22\\ (0.01)\end{tabular} & \begin{tabular}[c]{@{}c@{}}1.32\\ (0.01)\end{tabular} & \begin{tabular}[c]{@{}c@{}}1.35\\ (0.01)\end{tabular} &  & \begin{tabular}[c]{@{}c@{}}1.69\\ (0.02)\end{tabular} & \begin{tabular}[c]{@{}c@{}}1.61\\ (0.02)\end{tabular} & \begin{tabular}[c]{@{}c@{}}1.91\\ (0.06)\end{tabular} &  & \begin{tabular}[c]{@{}c@{}}19348.32\\ (56.11)\end{tabular} & \begin{tabular}[c]{@{}c@{}}19828.85\\ (49.92)\end{tabular} & \begin{tabular}[c]{@{}c@{}}19845.22\\ (89.54)\end{tabular} &  & \begin{tabular}[c]{@{}c@{}}194.07\\ (42.48)\end{tabular} & \begin{tabular}[c]{@{}c@{}}4243.31\\ (1084.48)\end{tabular} & \begin{tabular}[c]{@{}c@{}}2371.33\\ (1199.92)\end{tabular} \\ \cline{3-5} \cline{7-9} \cline{11-13} \cline{15-17} 
		(b) &  & \begin{tabular}[c]{@{}c@{}}1.32\\ (0.01)\end{tabular} & \begin{tabular}[c]{@{}c@{}}1.39\\ (0.01)\end{tabular} & \begin{tabular}[c]{@{}c@{}}1.43\\ (0.01)\end{tabular} &  & \begin{tabular}[c]{@{}c@{}}1.43\\ (0.03)\end{tabular} & \begin{tabular}[c]{@{}c@{}}1.47\\ (0.02)\end{tabular} & \begin{tabular}[c]{@{}c@{}}1.53\\ (0.02)\end{tabular} &  & \begin{tabular}[c]{@{}c@{}}23942.34\\ (90.71)\end{tabular} & \begin{tabular}[c]{@{}c@{}}24270.27\\ (76.68)\end{tabular} & \begin{tabular}[c]{@{}c@{}}24335.16\\ (88.15)\end{tabular} &  & \begin{tabular}[c]{@{}c@{}}150.03\\ (52.12)\end{tabular} & \begin{tabular}[c]{@{}c@{}}4548.67\\ (872.37)\end{tabular} & \begin{tabular}[c]{@{}c@{}}3127.96\\ (522.47)\end{tabular} \\ \cline{3-5} \cline{7-9} \cline{11-13} \cline{15-17} 
	\end{tabular}
}
\end{table}

\subsection{Estimation of the model components. Prediction of response trajectories}

We fit the model (\ref{SOW}) to the entire data. First, we examine the residuals: the auto-correlation (ACF) and the partial ACF (PACF) plots in Section B1 of the Supplementary Material show no evidence of auto-regressive dependence. In fact in Section B1 we investigated further dependence of the current feed intake onto the previous days feed intake and found no evidence of lagged auto-regressive dependence. Section B3 considers diagnostic plots for the other random components assumed by our model. Fitting model (\ref{SOW}) yields the following estimates of the random effects: $\widehat \sigma_0= 0.71$, which quantifies the variability of the mean feed intake (intercept) across sows, $\widehat \sigma_1 = 0.80$, which measures the variability of the sows rate of change of feed intake (slope), and $\widehat \sigma_g = 0.21$, which indicates the amount of variability of the group-level mean feed intake.

\begin{figure}[H]\centering
	
	\begin{tabular}{cc}
		\hspace{-.9cm} 
		\includegraphics[angle=0, height=6cm, width=7cm]{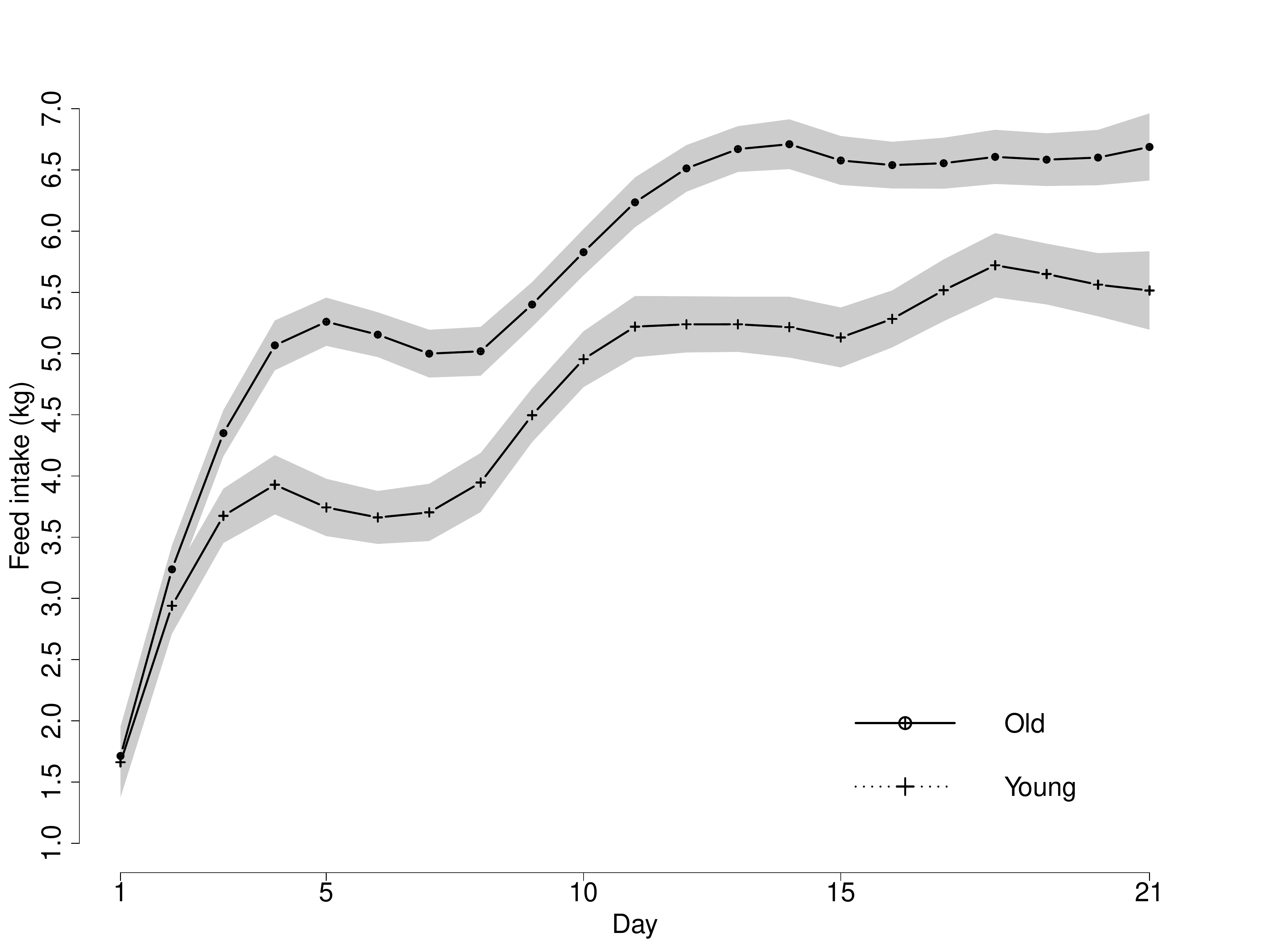}&
		\hspace{-0.75cm}\includegraphics[angle=0, height=6cm, width=8.5cm]{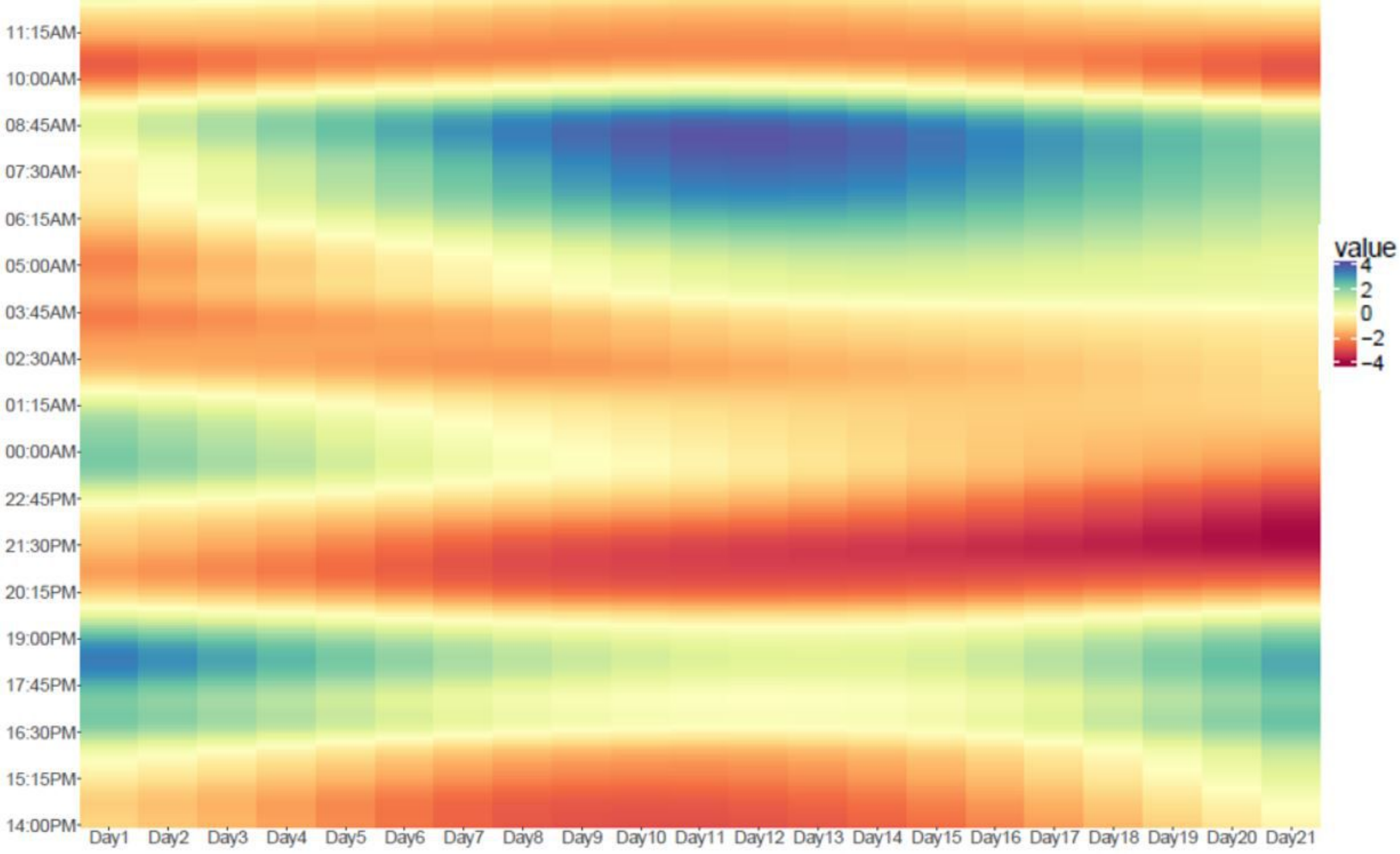}
		%\vspace{-0.5cm}
	\end{tabular}
	\caption{\small {Parameter estimates in the lactation sow application. Left panel depicts the estimated intercept function for the old (solid line) and young (dashed) sows with 95\% pointwise confidence intervals. Right panel shows the estimated regression coefficient $\widehat \gamma(\cdot, t)$ for each lactation day $t=1,2,\ldots, 21$. 
			\label{fig: lactatingsow-est}}}
\end{figure}

Figure \ref{fig: lactatingsow-est} shows the estimates of the model parameter functions. Specifically the left panel depicts the estimated mean feed intake for old sows (solid line), $\widehat \beta_1(t)$, and for young sows (dashed line), $\widehat \beta_0(t)$ along with 95\% pointwise confidence bands. It appears that the feed intake is about the same in the first couple of days, for both young and old sows, but shortly afterwards the older sows eat more than their younger relatives. By the fourth lactation day the older sows have an advantage of feed intake of up to 1-1.5 kg per day and they maintain this advantage for the remaining duration of the lactation period. 

Figure \ref{fig: lactatingsow-est}, in the right panel, shows the estimated regression coefficient $\widehat \gamma(s, t)$, which quantifies the minute-by-minute ($s$) effect of the {temperature} on the feed intake during the first 21st lactation days ($t$). As this regression coefficient function is identifiable only up to a function of $s$, we focus mainly on the changes across $t$. The association between the temperature and the feed intake changes during the lactation duration. For example, lower temperature levels around 7:30AM - 8:30AM are associated with much lower feed intake during the middle of the lactation period, say days 8 through 16, relative to the feed intake at the beginning or end of the lactation period. Also higher early evening (5:45PM - 7PM) temperature levels are associated with larger feed intake at the beginning of the lactation period, compared to the feed intake later on. These findings are based on the point estimates solely and do not account for uncertainty in the estimation; thus should be interpreted with caution. Additionally, using a version of $R-$square for this setting, the proposed model accounts for about $33\%$ of the variation in the data and the signal to noise ratio is estimated to be about $0.1$. Sections B.1-B.3 of the Supplementary Material include additional results for the data analysis, while Section B.4 provides prediction results for LDFR approach when a covariance model based on random group and subject random intercept is used. \\

\begin{figure}[H] %\centering
	\noindent\makebox[\textwidth]{
		\begin{tabular}{ccc}
			
			\includegraphics[angle=0, height=5cm, width=6.1cm]{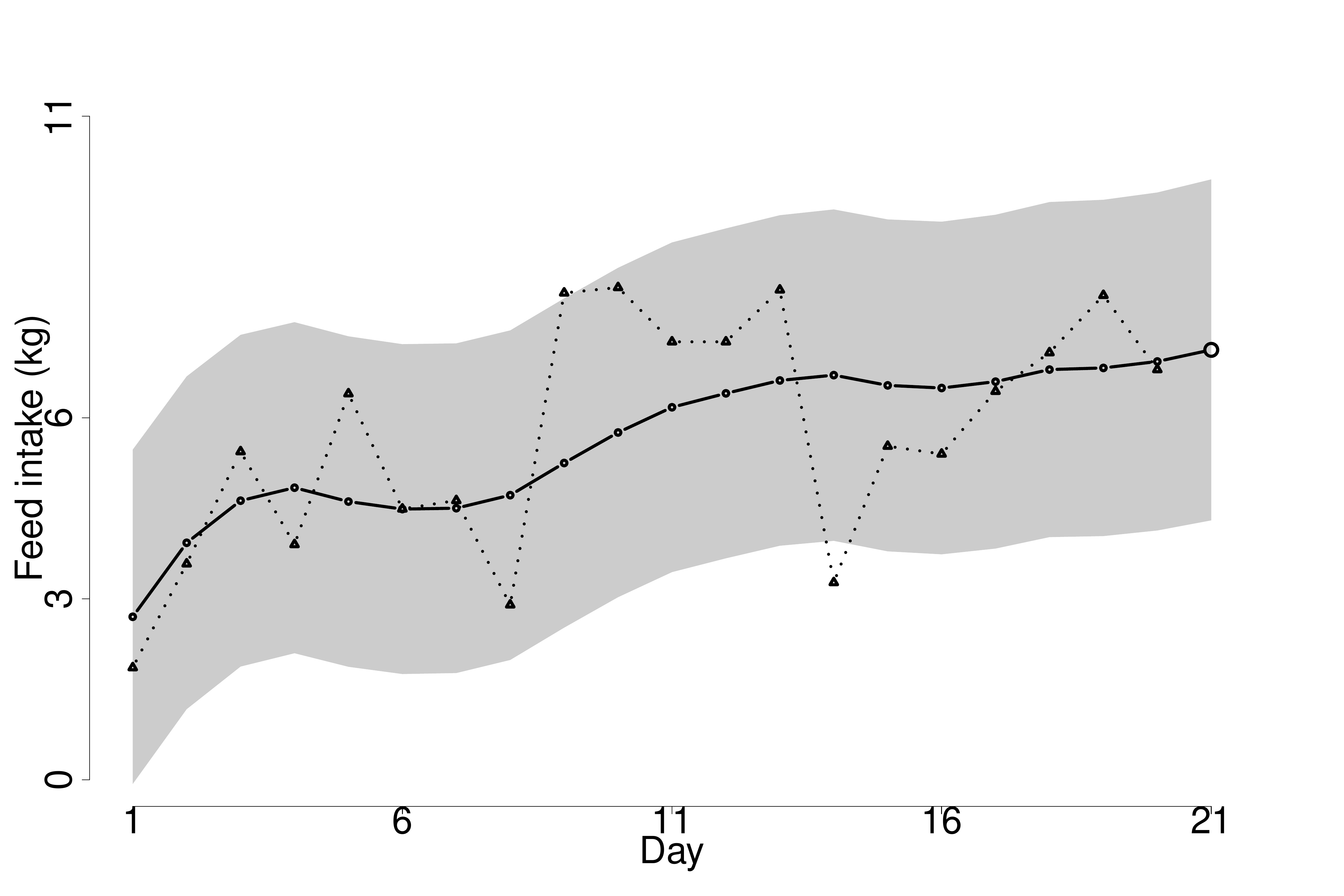}&

			\includegraphics[angle=0, height=5cm, width=6.1cm]{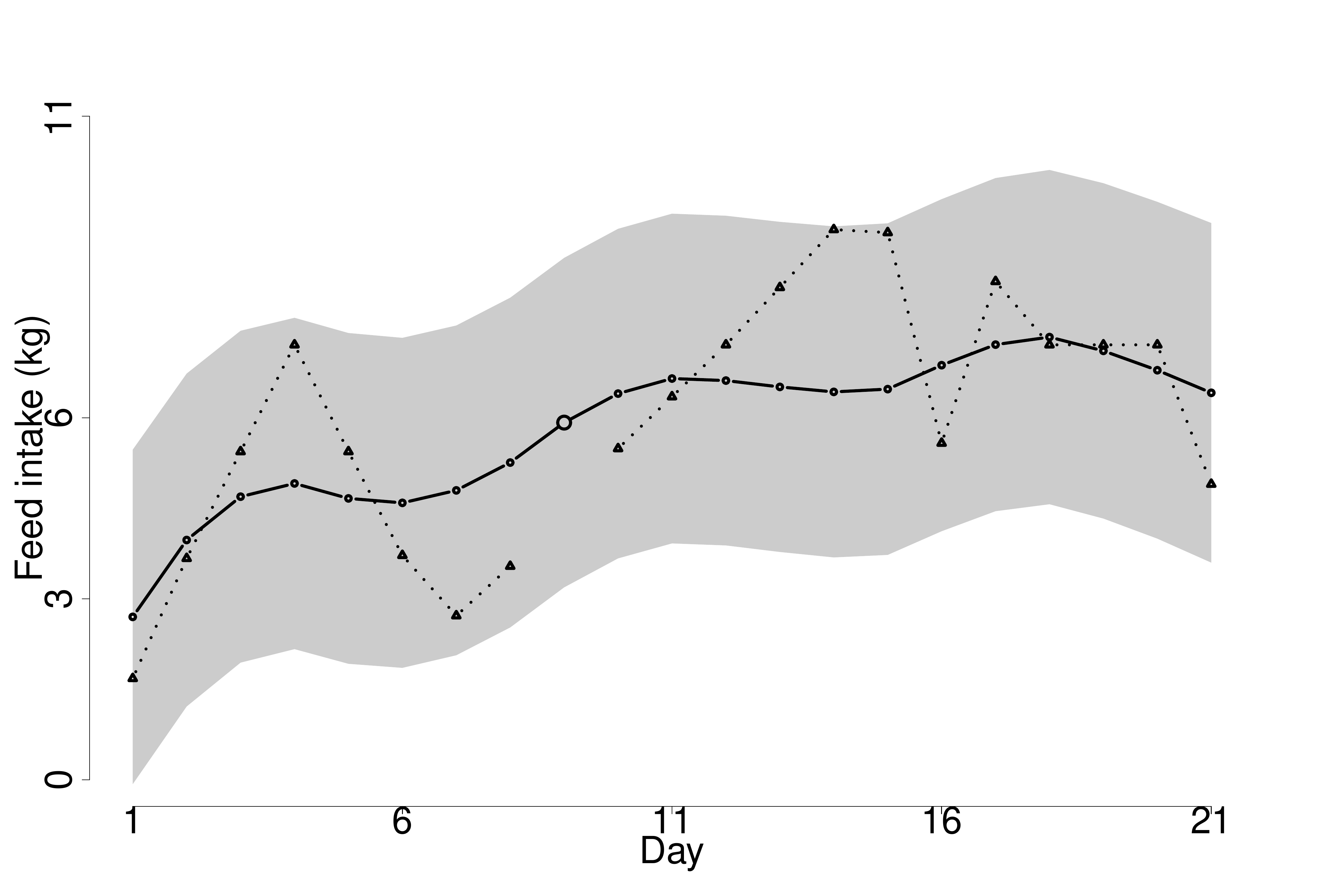}  & %\vspace{-.7cm}\\

			\includegraphics[angle=0, height=5cm, width=6.1cm]{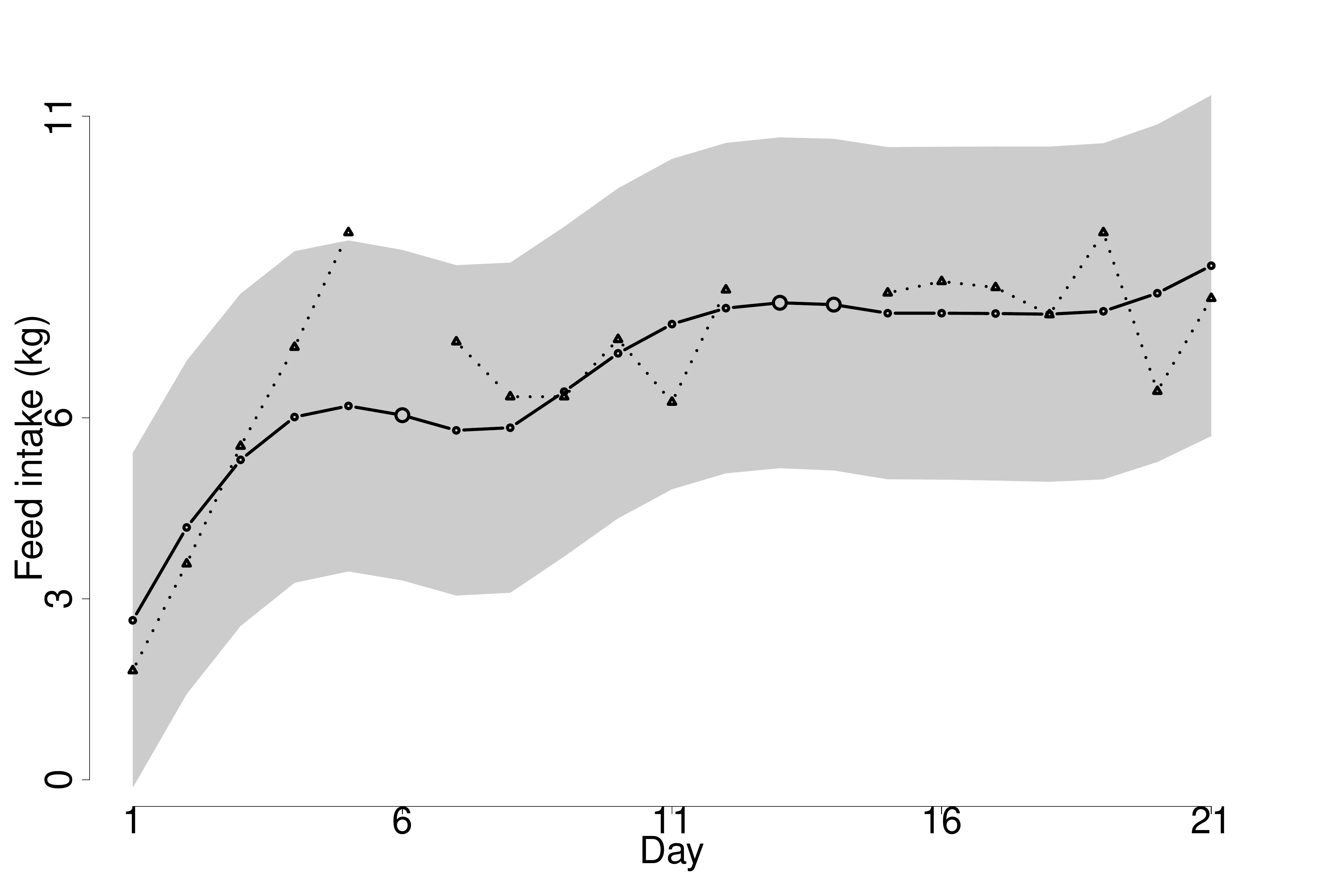}
			%\includegraphics[angle=0, height=5.2cm, width=6.1cm]{sow4.pdf}
			
			%\vspace{-0.5cm}
		\end{tabular}
	}
	\caption{\small Predicted full feed intake trajectories for two young (left and middle) and one old (right) sows. Shaded regions correspond to 95\% pointwise prediction bands based on LDFR.   
			\label{fig: lactatingsow-pred}}
\end{figure}

Figure \ref{fig: lactatingsow-pred} shows the predicted full trajectories of the feed intake for two young sows (left and middle panels) and one old sow (rightmost) selected at random from different groups along with their pointwise prediction bands constructed as detailed in Section \ref{sec:prediction}. The predicted trajectories are obtained from $\widehat FI_{i}(t) =   \widehat \beta_{p_{i}}(t) + \int \widehat Temp_{i}(s, t) \widehat \gamma(s,t) ds  + \widehat b_{g_i} + \widehat b_{0i(g_i)} + \widehat b_{1i(g_i)}t $ for every day $t=1,\ldots, 21$, where $\widehat Temp_{i}(\cdot , t)$ are the smooth and demeaned {temperature} profiles observed in relation to sow $i$, and $\widehat b_{g_i}, \widehat b_{0i(g_i)}$ and  $\widehat b_{1i(g_i)}$ are predicted effects. These results indicate too greater feed intake for older sows relative to their younger counterparts; the prediction intervals are wider to account for estimated measurement error.

%Full trajectory of $FI$ for four (two from each parity) randomly chosen sows.

\subsection{Validation of the results for data application via simulation study}

In this section we consider a simulation study mimicking the sow data structure. In particular, we generate feed-intake (kg) using the model (\ref{SOW}) with the estimated smooth effects $\widehat \beta_{p_{i}}(\cdot)$ and $\widehat \gamma(\cdot, \cdot).$ The dependence across measurements are modeled through $b_{g_i} + b_{0i(g_i)} + b_{1i(g_i)}t_{ij} + \varepsilon_{ij}$ resembling the covariance structure of the data; here all terms bear the usual meaning as before and are generated as $b_{g_i} \sim \pN (0, \widehat \sigma_{g}^{2}),$ $b_{0i(g_i)} \sim \pN (0, \widehat \sigma_{0}^{2}),$ $b_{1i(g_i)} \sim \pN (0, \widehat \sigma_{1}^{2}),$ and $\varepsilon_{ij} \sim \pN(0, \widehat \sigma_{e}^{2}).$ Temperature profiles are constructed as $Temp_{ij}(s) = \widehat \tau(s , t_{ij}) + \sum_{k=1}^{7} \xi_{ik}(t_{ij}) \widehat \phi_{k}(s)$ where the mean $\widehat \tau(s , t_{ij})$ and the fPCs $\{\widehat \phi_{k}(\cdot); k = 1, \cdots, 7\}$ are obtained from the data, and the scores $\xi_{ik}(\cdot)$ are generated as a zero-mean random process with covariance $\widehat G_{k}(\cdot, \cdot)$ for each $k$; see Section B.3 of the Supplementary Material. Denote the observed temperature profiles by $nTemp_{ij}(s) = Temp_{ij}(s) + \epsilon_{1,ij}(s) + \epsilon_{2,ij}(s);$ where $\epsilon_{1,ij}(s)$ is a smooth error process with zero mean and covariance $\widehat \Gamma(s,s'),$ and $\epsilon_{2,ij}(s)$ is a white noise with zero-mean and covariance $\widehat \sigma_{w}^{2}\mathds{1}(s=s').$  We consider $I = 475,$ $m_{i} = \{7, \cdots, 21\},$ and $g_{i} = 1, \cdots, 21$ as same as that of data. We simulate the data for 100 times, and split each dataset into a training and test set on which prediction performance is evaluated. We also assess the prediction coverage for both existing ($i$) and new ($i^{*}$) sows feed-intake. For the latter case, we construct a new set of $125$ sows and for each of them we simulate temperature profiles according to our model; the data for these 125 sows is not used in the estimation of the model parameters. 

{Table \ref{sim_result_sow} compare the prediction performance of the three approaches. LDFR and LPEER show similar accuracy, while LPFR remains inferior. However LDFR exhibits better fitting criteria as assessed by AIC. In addition, the average prediction coverage stays around the nominal levels (i.e. 95\% and 90\%) for both the observed and unobserved sows while having larger prediction band width for the unobserved ones. These findings are in agreement with the results demonstrated in Table \ref{Gaussian_coverage} and \ref{comparison_sow_temperature}.}

\begin{table}[H]
	\tiny
\centering
\caption{Numerical results based on 100 Monte Carlo simulations mimicking sow data. Reported are the median prediction errors, marginal AICs, RMPE$_{trj},$ IQR (in parenthesis), average coverage probabilities of 95\% and 90\% pointwise prediction bands, standard errors (in parenthesis with superscript $\dagger$), and average length of intervals (in square bracket) for the existing ($Y_{i}$) and new sows ($Y_{i^{*}}$).}
\label{sim_result_sow}
\noindent\makebox[\textwidth]{
	\begin{tabular}{ccccccccccccc}
		& \multicolumn{2}{c}{IN$_{PE}$} & \multicolumn{2}{c}{OUT$_{PE}$} & \multicolumn{2}{c}{AIC} & \multicolumn{2}{c}{RMPE$_{trj}$} & \multicolumn{2}{c}{Cov$^{0.95}_{trj}$} & \multicolumn{2}{c}{Cov$^{0.90}_{trj}$} \\ \cline{2-13} 
		& $(a)$ & $(b)$ & $(a)$ & $(b)$ & $(a)$ & $(b)$ & $Y_{i}$ & $Y_{i^{*}}$ & $Y_{i}$ & $Y_{i^{*}}$ & $Y_{i}$ & $Y_{i^{*}}$ \\ \cline{2-13} 
		LDFR & \begin{tabular}[c]{@{}c@{}}1.29\\ (0.02)\end{tabular} & \begin{tabular}[c]{@{}c@{}}1.31\\ (0.02)\end{tabular} & \begin{tabular}[c]{@{}c@{}}1.53\\ (0.03)\end{tabular} & \begin{tabular}[c]{@{}c@{}}1.43\\ (0.03)\end{tabular} & \begin{tabular}[c]{@{}c@{}}19669.47\\ (257.97)\end{tabular} & \begin{tabular}[c]{@{}c@{}}23899.41\\ (344.42)\end{tabular} & \begin{tabular}[c]{@{}c@{}}1.33\\ (0.01)\end{tabular} & \begin{tabular}[c]{@{}c@{}}1.61\\ (0.05)\end{tabular} & \begin{tabular}[c]{@{}c@{}}0.96 (0.02)$^{\dagger}$\\ {[}5.50{]}\end{tabular} & \begin{tabular}[c]{@{}c@{}}0.95 (0.02)$^{\dagger}$\\ {[}6.32{]}\end{tabular} & \begin{tabular}[c]{@{}c@{}}0.92 (0.03)$^{\dagger}$\\ {[}4.62{]}\end{tabular} & \begin{tabular}[c]{@{}c@{}}0.90 (0.03)$^{\dagger}$\\ {[}5.30{]}\end{tabular} \\ \cline{2-13} 
		LPEER & \begin{tabular}[c]{@{}c@{}}1.35\\ (0.02)\end{tabular} & \begin{tabular}[c]{@{}c@{}}1.38\\ (0.02)\end{tabular} & \begin{tabular}[c]{@{}c@{}}1.56\\ (0.03)\end{tabular} & \begin{tabular}[c]{@{}c@{}}1.47\\ (0.03)\end{tabular} & \begin{tabular}[c]{@{}c@{}}19999.92\\ (258.97)\end{tabular} & \begin{tabular}[c]{@{}c@{}}24207.91\\ (323.01)\end{tabular} & NA & NA & NA & NA & NA & NA \\ \cline{2-13} 
		LPFR & \begin{tabular}[c]{@{}c@{}}1.42\\ (0.02)\end{tabular} & \begin{tabular}[c]{@{}c@{}}1.42\\ (0.02)\end{tabular} & \begin{tabular}[c]{@{}c@{}}1.69\\ (0.05)\end{tabular} & \begin{tabular}[c]{@{}c@{}}1.53\\ (0.03)\end{tabular} & \begin{tabular}[c]{@{}c@{}}20138.64\\ (283.82)\end{tabular} & \begin{tabular}[c]{@{}c@{}}24262.62\\ (328.75)\end{tabular} & NA & NA & NA & NA & NA & NA \\ \cline{2-13} 
	\end{tabular}
}
\end{table}

\section{Discussion} \label{sec:discussion}

In this paper we consider longitudinal dynamic functional regression for scalar responses and functional covariates observed in a longitudinal design. We propose a flexible way to model the time-varying bivariate regression coefficient function by combining ideas from functional data analysis and longitudinal data analysis. As one anonymous reviewer asserted, this clever combination allows one to tackle a challenging problem that has not previously been solved in this generality. The methodology relies on the assumptions that the leading eigenbasis functions of the functional predictor are most predictive of the response and that the latent predictor signals are relatively  smooth. The approach is applicable to Gaussian as well as non-Gaussian responses and can directly accommodate additional vector covariates, non-linear effects of vector covariates, as well as multiple functional covariates observed on diverse sampling designs and with measurement error. The methodology can be easily implemented using the existing freely available software.

Numerical results show that the prediction performance of our approach is superior to existing alternative approaches when the regression coefficient function is indeed time varying, and is very competitive with the existing alternatives when the regression coefficient function is time-invariant. In spite of the increased flexibility, this method is computationally efficient; in fact it is orders of magnitude faster than its closest competitor. We discuss an approach to reconstruct the full response trajectory. We applied the method to the animal science application and found that the effect of the temperature on the feed intake of the lactating sows varies with the days since they gave birth.

One limitation of our methodology is that it relies on the implicit assumption that the current response is related to the current functional predictor only i.e. $E[Y_{ij}|X_{i1}(\cdot), \cdots, X_{in_{i}}(\cdot)] = E[Y_{ij}|X_{ij}(\cdot)].$ While this assumption makes sense for our application, it may not be reasonable for other situations. One possible approach to account for the past functional covariates is by considering a regression model inspired by the historical functional linear model (see \cite{malfait2003historical,scheipl2015functional,pomann2016lag} and \cite{kim2011recent}).

\section*{Supplementary Material}

Additional simulation results as well as additional data analysis results are presented as Supplementary Material. Moreover, the R-code for implementation of the proposed framework is posted publicly at
\url{http://www4.stat.ncsu.edu/~staicu/software/LDFR.zip}. The fitting methodology is illustrated at \url{http://www4.stat.ncsu.edu/~staicu/software/illustration_LDFR.html} using a generated data set.

\section*{Acknowledgment}
Staicu's research was funded by National Science Foundation grant DMS 1454942 and National Institute of Health grants R01 NS085211 and R01 MH086633.  The data used originated from work supported in part by the North Carolina Agricultural Foundation, Raleigh, NC.

\bibliographystyle{apalike}
\small
\bibliography{LDFR2}

%\newpage

\end{document}